\documentclass[usenatbib]{mn2e}
\usepackage{graphicx}
\usepackage{rotating} 

\newcommand{\lsim }{{\lower0.8ex\hbox{$\buildrel <\over\sim$}}}
\newcommand{\gsim }{{\lower0.8ex\hbox{$\buildrel >\over\sim$}}}

\usepackage{soul}

\def\Chandra{\emph{Chandra}}
\def\Chandraacis{\emph{Chandra/ACIS}}

\def\HST{${\it HST}$}
\def\XMM{{\it XMM-Newton}}

\def\simge{\mathrel{%
  \rlap{\raise 0.511ex \hbox{$>$}}{\lower 0.511ex \hbox{$\sim$}}}}
\def\simle{\mathrel{
  \rlap{\raise 0.511ex \hbox{$<$}}{\lower 0.511ex \hbox{$\sim$}}}}

\newcommand{\Msun}{\ifmmode {M_{\odot}}\else${M_{\odot}}$\fi}
\newcommand{\Lsun}{\ifmmode {L_{\odot}}\else${L_{\odot}}$\fi}
\newcommand{\Rsun}{\ifmmode {R_{\odot}}\else${R_{\odot}}$\fi}

\usepackage{xargs}
\usepackage{color}

\begin{document}

\title[Mass/Radius Constraints for Two Neutron Stars]{Improved Mass and Radius Constraints for Quiescent Neutron Stars in $\omega$ Cen and NGC 6397 
}

\author[Heinke et al.]{ C.~O.~Heinke$^{1}$\thanks{heinke@ualberta.ca}, H.~N.~Cohn$^{2}$, P.~M.~Lugger$^{2}$, 
N.~A.~Webb$^{3}$, 
W.~C.~G.~Ho$^4$, \newauthor
J.~Anderson$^5$, 
S.~Campana$^6$, 
S.~Bogdanov$^7$, 
D.~Haggard$^8$, 
A.~M.~Cool$^9$, J.~E.~Grindlay$^{10}$ \smallskip\\
$^{1}${Dept. of Physics, University of Alberta, CCIS 4-183, Edmonton, AB T6G 2E1, Canada}\\
$^{2}${Department of Astronomy, Indiana University, 727 E. Third St, Bloomington, IN 47405, USA}\\
$^3${Universite de Toulouse, UPS-OMP, IRAP, Toulouse, France; CNRS, IRAP, 9 Av. Colonel Roche, BP 44346, F-31028 Toulouse, Cedex 4, France}\\
$^4${Mathematical Sciences \& STAG Research Centre, University of Southampton, Southampton SO17 1BJ, UK}\\
$^5${Space Telescope Science Institute, 3700 San Martin Drive, Baltimore, MD 21218, USA}\\
$^6${INAF - Osservatorio astronomico di Brera, Via Bianchi 46, Merate I-23807 (LC), Italy}\\
$^7${Columbia Astrophysics Laboratory, Columbia University, 550 West 120th Street, New York, NY 10027, USA}\\
$^8${CIERA, Physics and Astronomy Department, Northwestern University, 2145 Sheridan Road, Evanston, IL 60208, USA; CIERA Fellow}\\
$^9${Department of Physics and Astronomy, San Francisco State University, 1600 Holloway Avenue, San Francisco, CA 94132, USA}\\
$^{10}${Harvard-Smithsonian Center for Astrophysics, 60 Garden Street, Cambridge, MA 02138, USA}\\
}
\maketitle

\begin{abstract}
We use \Chandra\ and \XMM\ observations of the globular clusters $\omega$ Cen and NGC 6397 to measure the spectrum of their quiescent neutron stars (NSs), and thus to constrain the allowed ranges of mass and radius for each.  We also use {\it Hubble Space Telescope} photometry of NGC 6397 to identify a potential optical companion to the quiescent NS, and find evidence that the companion lacks hydrogen.  We carefully consider a number of systematic problems, and show that the choices of atmospheric composition, interstellar medium abundances, and cluster distances can have important effects on the inferred NS mass and radius.  We find that for typical NS masses, the radii of both NSs are consistent with the $10-13$ km range favored by recent nuclear physics experiments. This removes the evidence suggested by Guillot and collaborators for an unusually small NS radius, which relied upon the small inferred radius of the NGC 6397 NS.
\end{abstract}

\begin{keywords}
Globular clusters -- X-rays: binaries -- stars: neutron -- dense matter
\end{keywords}

\section{Introduction}
The behaviour of matter at extremely high densities cannot be measured in laboratories on Earth, but can be probed through measurements of the properties of neutron stars (NSs).  Some NS properties can be measured accurately without systematic uncertainties, such as their rotation rates, velocities, and, for some, their gravitational masses.  However, effective constraints on the interior makeup of NSs (and thus, their equation of state) require robust measurements of both mass and radius, ideally for several NSs \citep{Lattimer07}.  
Recent attention has focused on thermal surface radiation from NSs, during thermonuclear bursts and/or during periods of quiescence \citep{Ozel10,Guillot13}.  Both the apparent surface area and the NS's Eddington limit can be tested during observations of thermonuclear bursts, thus providing orthogonal constraints, but a number of systematic uncertainties (e.g. anisotropies, variations in the persistent emission, the radius of emission, the details of the spectrum) remain unresolved \citep[e.g.][]{Steiner10,Zamfir12,Galloway12,Worpel13}.  

Fewer systematic uncertainties surround measurements of the radiation radius of quiescent NSs, or quiescent low-mass X-ray binaries (qLMXBs) in globular clusters \citep{Rutledge02a}. The distances to globular clusters are well-known, with a 6\% rms variation among different distance measurement techniques \citep{Woodley12}.  NSs in qLMXBs produce thermal (blackbody-like) X-rays due to heating of the NS core (and crust) during periods of accretion \citep{Brown98,Rutledge02b}, and seem to produce both thermal X-rays and nonthermal X-rays (typically fit with a power-law) by accretion, though other physics may be involved in the poorly-understood power-law  \citep{Campana98a,Cackett10,Deufel01}.  
NSs in qLMXBs are believed to have low magnetic fields, and in most cases pure hydrogen atmospheres, allowing for robust physical modeling \citep{Zavlin96,Rajagopal96,Heinke06a,Haakonsen12}.  The constraints in mass and radius from these measurements are degenerate along a curved track, close to a line of constant $R_{\infty}=R/\sqrt{1-2GM/(R c^2)}$, where $M$ and $R$ are the NS mass and radius, and $R_{\infty}$ is the radius as seen at infinity  \citep[e.g. ][]{Heinke06a}.  This is due to gravitational redshift increasing with increasing mass, shifting the intrinsic NS temperature to higher values, and thus requiring a smaller emitting surface area.

Numerous qLMXBs are known in globular clusters \citep[e.g.][]{Heinke03d,Guillot09a}, of which a handful are sufficiently bright and well-observed to provide interesting constraints on mass and radius (e.g., in 47 Tuc, \citealt{Heinke06a}; in NGC 6397, \citealt{Guillot11}; in $\omega$ Cen and M13, \citealt{Webb07}; in M28, \citealt{Servillat12}).
However, radius measurements of NSs in globular cluster qLMXBs reveal significant discrepancies, with the inferred radii for some NSs appearing significantly larger or smaller than the inferred radii for other NSs (e.g. \citealt{Webb07}).  We particularly note the Bayesian multi-object analysis of \citet{Guillot13}, who conducted a simultaneous spectral analysis of five qLMXBs in globular clusters to measure the NS radius. 
They prefer an extremely small radius for the qLMXB  U24 in NGC 6397 ($R_{\infty}$=$8.4^{+1.3}_{-1.1}$ km, or $R$=6.6$^{+1.2}_{-1.1}$ km), and an extremely large radius or mass for the $\omega$ Cen qLMXB ($R_{\infty}$=23.6$^{+7.6}_{-7.1}$ km, or $R$=20.1$^{+7.4}_{-7.2}$ km).  Although a single NS equation of state can pass through both of these mass-radius regions (at nearly constant radius in the relevant mass range), this would generally require unphysically low masses ($<$1 \Msun) for U24, and rather large masses ($>$2 \Msun) for the $\omega$ Cen NS.  

This work is motivated by a desire to see if these NS measurements can be reconciled. We are also interested in investigating the effect of systematic uncertainties in the chemical composition of the NS atmosphere and in the modeling of the interstellar medium.  We utilize our new deep \Chandra\ observations of $\omega$ Cen, along with archival \Chandra\ and \XMM\ observations, and our archived \Chandra\ observations of NGC 6397.  We also use results from our deep {\it Hubble Space Telescope} (\HST) observations of NGC 6397 to constrain the nature of the companion to U24 in NGC 6397.  Before describing the data and our analysis, we discuss some systematic problems we consider here (not an exhaustive list; see e.g. \citealt{Guillot11}, \citealt{alGendy14}, and finally Elshamouty et al.\ in prep., for limits on, and detailed consideration of the effects of, temperature inhomogeneities).  

\subsection{Systematics: Atmosphere Composition}

Ionized hydrogen atmospheres (for low magnetic fields and temperatures relevant to our problem, $>3\times10^5$ K, \citealt{Zavlin96}, the atmosphere is completely ionized) alter the outgoing flux from the surface through free-free absorption.  The opacity dependence of this absorption is roughly $\nu^{-3}$, meaning that the observed flux is shifted to higher energies, while appearing roughly blackbody in shape \citep{Romani87,Zavlin96,Rajagopal96}.  The timescale of only a few seconds for elements to stratify in a NS atmosphere, means that the lowest-density element will rise to the surface \citep{Alcock80,Hameury83,Brown02}, unless heavier elements are being deposited at a high rate, translating to an accretion luminosity of roughly $10^{33}$ ergs/s \citep{Rutledge02a}. 
Such an accretion rate may be experienced by, e.g., Aquila X-1 in some quiescent observations, based on its variability properties. Aquila X-1 may also show evidence of features in its spectrum from heavier elements \citep{Rutledge02a}.  
 If the accreted material possesses hydrogen, and the luminosity is well below $10^{33}$ ergs/s (or shows a total lack of evidence of accretion, \citealt{Heinke06a}) the emitted spectrum of the NS should be well-represented by a hydrogen atmosphere model.  The hydrogen atmosphere models constructed by different groups give reproducible results \citep{Zavlin96,Lloyd03,Heinke06a,Haakonsen12}, indicating that parameters derived from hydrogen atmosphere modeling should be accurate to within a few percent \citep{Haakonsen12}.

However, the accreted material may not possess any hydrogen, if the donor star is a white dwarf.  It is possible that the accreted material may spallate on impact, producing protons \citep{Bildsten93}, but spallation may require infalling protons that would not be present if the donor star is a white dwarf \citep{intZand05}.  Also, the creation of hydrogen would alter the character of observed thermonuclear bursts from NSs accreting from white dwarfs, which show distinct properties consistent with the absence of hydrogen \citep[e.g.][]{Cumming03}.
White dwarf donors have been confirmed (by orbital period measurement) for five luminous (persistent or transient, reaching $L_X>10^{35}$ ergs/s) NS X-ray binaries in globular clusters, out of 18 such sources (of which 10 are confirmed not to have white dwarf donors; \citealt{Zurek09,Altamirano10,Bahramian13}).  

Thus, there is good reason to expect the presence of qLMXBs in globular clusters with helium (or possibly carbon) atmospheres.  Helium atmospheres generally resemble hydrogen atmospheres \citep{Romani87}, but are slightly harder--the difference between the color temperature and the effective surface temperature is slightly larger than for hydrogen atmospheres \citep{Ho09}. Helium atmosphere models will thus predict different best-fit values of mass and radius for observed spectra.  (Carbon atmospheres are significantly harder, and show a strong edge near 0.3 keV;  \citealt{Ho09}.  We do not consider carbon atmospheres further in this work, as they seem inapplicable to the qLMXBs of interest; fitting a carbon atmosphere would give much larger inferred radii.)  We have applied the helium models of \citet{Ho09} to the high-quality X-ray spectra of qLMXBs in M28 \citep{Servillat12} and M13 \citep{Catuneanu13}, finding significantly larger inferred values of $R_{\infty}$ (and thus, larger masses or radii) for helium than hydrogen models.  
This motivates us to consider fitting a helium atmosphere model to U24 in NGC 6397, as its inferred radius, for a hydrogen atmosphere, is rather small \citep{Guillot13,Lattimer14}.  Note that the donor to the qLMXB in $\omega$ Cen is known to possess hydrogen, through detection of H$\alpha$ in emission \citep{Haggard04}.

\subsection{Systematics: Interstellar Medium}

Extinction by the interstellar medium (ISM) significantly affects the inferred radius of the NSs \citep{Lattimer14}. Thus, careful modeling of the ISM is critical.  The absorption of X-rays is principally due not to hydrogen, but to the heavier elements in the ISM, so their relative abundances will affect the shape of the absorbed flux. Current modeling of the abundances and cross-sections of the ISM \citep[][incorporated as the {\tt tbabs} model in XSPEC]{Wilms00} show significant differences with older models such as \citet{Morrison83}, incorporated as the {\tt wabs} model used by \citet{Guillot13}.  We will test whether the choice of absorbing model makes a difference here.

\citet{Lattimer14} raise the question of whether the interstellar absorption column, $N_H$, for each qLMXB in a globular cluster should be fixed to the value derived from an independent HI survey (such as \citealt{Dickey90}) in the direction of the cluster, arguing that fixing $N_H$ in this manner gives more consistent $R_{\infty}$ values for different NSs, and that the locus of $R_{\infty}$ measurements is more consistent with predictions of the equation of state from nuclear experiments and theoretical neutron matter studies. Fixing $N_H$ to values from an HI survey has clear flaws, since the HI surveys account only for atomic H, include all H along the line of sight (some of which may lie behind the cluster), and are integrated over large angular scales (possibly not representative of the particular line of sight).  Optical measurements of the extinction towards the globular cluster are superior to HI surveys (as they avoid these problems), but may suffer from uncertainty in the transformation from $A_V$ to $N_H$ (based, e.g., on uncertain abundances), and do not allow for the possibility of extra $N_H$ intrinsic to the binary system. 
Spectral fitting of X-ray data gives a direct measurement of the $N_H$ experienced along the line of sight, and should be used if accurate constraints on NS radii are desired.  
For our analyses in this paper, we measure the $N_H$ directly from the qLMXB X-ray spectra without any external constraints.  We find that these measurements are typically close to the $N_H$ values calculated from the measured $A_V$ to each cluster (and are often closer to these values than to the $N_H$ inferred from independent HI surveys), but that in some cases they are significantly different.

\subsection{Systematics: Distance to Globular Clusters}

The distances to globular clusters are a topic of great importance for stellar evolution and cosmology studies \citep{Krauss03}. No method of measuring distances to globular clusters is completely without systematic uncertainties.  \citet{Guillot13} prefer dynamical methods (comparing radial velocities to proper motions) for measuring distances, using the argument that the uncertainties in these methods are well-understood.  However, this understates the systematic uncertainties in dynamical distance derivations, coming from the use of different stars for the radial velocities and the proper motions, and from the assumption of isotropy in velocities.  

For $\omega$ Cen, the dynamical distance estimate by \citet{vandeVen06} is based on thorough dynamical modeling of an inclined, axisymmetric, rotating ellipsoid, giving a distance of 4.8$\pm$0.3 (1$\sigma$) kpc.  This measurement may be slightly underestimated if some interloping stars remain in the proper motion data, and particularly by the systematic offset between the proper motions of the metal-rich RGB stars and the metal-poor RGB and HB populations \citep{Platais03,vandeVen06}.  Distance determinations using modeling of eclipsing binaries (5.36$\pm$0.3 kpc, \citealt{Thompson01}), RR Lyrae near-IR photometry (5.5$\pm$0.04 kpc, \citealt{delPrincipe06}), and the edge of the RR Lyrae instability strip (5.6$\pm$0.3 kpc, \citealt{Caputo02}) are larger, while SED modeling of giants gives 4.85$\pm$0.2 kpc \citep{McDonald09}.  The homogeneously calculated horizontal-branch distances of \citet{Harris96} (2010 revision)\footnote{http://physwww.physics.mcmaster.ca/$\sim$harris/mwgc.dat} give an $\omega$ Cen distance of 5.2 kpc. \citet{Bono08} measure the distance ratio between $\omega$ Cen and the better-studied cluster 47 Tuc with three different relative methods, all indicating that $\omega$ Cen is 16\% ($\pm$3\%) farther than 47 Tuc. \citet{Woodley12} average 22 published distance measurements for 47 Tuc to find d$_{47 Tuc}$=4.57 kpc (standard deviation of 0.28 kpc, error in mean of 0.06 kpc), in agreement with the best white dwarf distance measurement to 47 Tuc by \citet{Hansen13} of 4.61$\pm0.19$ kpc.
 Using this average distance for 47 Tuc, with its error in the mean, with the relative distance determination above, implies that d$_{\omega Cen}$=5.30$\pm0.17$ kpc.  
We choose for this paper to adopt 5.30$\pm0.17$ kpc as our standard 
for $\omega$ Cen's distance.

NGC 6397 has been the subject of recent high-quality distance measurements using the white dwarf cooling sequence (\citealt{Hansen07}, 2.54$\pm0.07$ kpc; \citealt{Strickler09}, 2.39$\pm0.13$ kpc using Hansen's WD mass estimate), and main-sequence fitting (\citealt{Gratton03}, 2.52$\pm0.10$ kpc).  The \citet{Harris96} (2010 revision) catalog calculates a horizontal branch distance of 2.3 kpc for NGC 6397.  
\citet{Guillot13} use a dynamical distance measurement of 2.02$\pm0.18$ kpc obtained by \citet{Rees96}.\footnote{The dynamical distance measurement of  \citet{Rees96} is not explained in detail, is not peer-reviewed (it is only listed in a conference proceeding), and is referred to as ``preliminary'' in a later conference proceeding by the same author \citet{Rees97}.}  \citet{Heyl12} performed a higher-quality dynamical distance measurement, using multiple deep \HST\ observations to obtain proper motions, obtaining a final measurement of $2.2^{+0.5}_{-0.7}$ kpc with minimal systematic errors, or a more precise measurement of $2.0\pm0.2$ kpc that is more vulnerable to systematic uncertainties (as the same stars are not used for radial velocities vs. proper motions). 
Dynamical distance methods generally assume no anisotropies among the stellar velocities, which could bias the calculated distance; such anisotropies have been measured by \citet{Richer13} in 47 Tuc.  
A relative distance comparison of NGC 6397 to 47 Tuc \citep{Hansen13} finds that NGC 6397 is 1.32$\pm0.10$ magnitudes closer, or 54.5$\pm$2.5\% of 47 Tuc's distance. Using the Woodley et al.\ 47 Tuc average distance measurement, we thus derive a distance of 2.51$\pm$0.07 kpc. 
Therefore, we choose a standard distance of 2.51$\pm$0.07 kpc, which aligns with the most accurate relative distance measurements, and is consistent (via relative distance measurements of each cluster to 47 Tuc) with our choice for $\omega$ Cen above.


\subsection{Systematics: Instrumental Calibration}

The absolute calibration of X-ray instruments, and particularly the cross-calibration between the \XMM\ pn and MOS, and \Chandra\ ACIS, detectors, is a topic of active research by the International Astronomical Consortium for High Energy Calibration (IACHEC) calibration consortium \footnote{http://web.mit.edu/iachec/papers/index.html}.  However, it has rarely been discussed in the literature on qLMXB radius measurements, with the exception of \citet{Catuneanu13}.  
Both the normalization of the X-ray flux, and the spectral shape measurements, show differences between different detectors.  \citet{Tsujimoto11}, comparing \Chandra/ACIS, \XMM\ pn and MOS, {\it Suzaku}/XIS and {\it Swift}/XRT spectra of the absorbed pulsar wind nebula G21.5-0.9 (1$-$8 keV), found that the \Chandra/ACIS detector measured fluxes 10\% higher than the average, while the \XMM/pn detector measured fluxes 10\% lower than the average among other detectors.  
\citet{Nevalainen10}, comparing \Chandra/ACIS and \XMM/EPIC observations of galaxy clusters, found that the 2$-$7 keV EPIC and ACIS fluxes were only 5$-$10\% different.  However,  measurement of cluster temperatures in the 0.5-2 keV band gave differences of 18\% between the \Chandra/ACIS and \XMM/pn, due to the ratio of ACIS data to the best pn model fit declining from 1.0 at 2 keV, to 0.9 at 0.5 keV.  Since qLMXB thermal spectra provide most of their counts between 0.5 and 2 keV, this suggests that \Chandra/ACIS and \XMM/pn will give diverging results at very high S/N ratios. We will investigate this question below for the $\omega$ Cen spectra, by including fits allowing the temperature, or a normalization constant between detectors, to vary.

\section{Observations}\label{sec:data}

\subsection{$\omega$ Cen: \Chandra, \XMM\ Data}\label{sec:wCen}

The available X-ray data on the $\omega$ Cen qLMXB (summarized in Table 1) include a 2001 \XMM\ EPIC observation \citep{Gendre03a}, two 2000 ACIS-I \Chandra\ observations totaling 69 ks \citep{Haggard09}, and two deeper 2012 ACIS-I \Chandra\ observations totaling 222 ks  \citep{Haggard13}.  

\begin{table*}  
\begin{center}
\caption{\bf \Chandra\ and \HST\ data used}
\begin{tabular}{lllll}
Telescope & Instr./filter & ID & Date   & Exp. time \\
\hline
\multicolumn{5}{c}{\bf $\omega$ Cen}\\
\hline
 \Chandra\   & ACIS-I & 653 & 2000 Jan. 24 & 25 ks \\
 \Chandra\   & ACIS-I & 1519 & 2000 Jan. 25 & 44 ks \\
\Chandra\   & ACIS-I & 13727 & 2012 Apr. 16 & 49 ks \\
 \Chandra\   & ACIS-I & 13726 & 2012 Apr. 17 & 174 ks \\
 \XMM\   & EPIC/medium & 112220101 & 2001 Aug. 12 & 40 ks \\
\hline
\multicolumn{5}{c}{\bf NGC 6397}\\
\hline
 \Chandra\   & ACIS-I & 79 & 2000 Jul. 31 & 48 ks \\
 \Chandra\   & ACIS-S & 2668 & 2002 May 13 & 28 ks \\
 \Chandra\   & ACIS-S & 2669 & 2002 May 15 & 27 ks \\
 \Chandra\   & ACIS-S & 7461 & 2007 Jun. 22 & 89 ks \\
 \Chandra\   & ACIS-S & 7460 & 2007 Jul. 16 & 148 ks \\
\hline
\HST\ & ACS/WFC/F435W & 10257 & 2004 Aug--2005 Jun & 1765 s\\
\HST\ & ACS/WFC/F625W & 10257 & 2004 Jul--2005 Jun & 1750 s\\
\HST\ & ACS/WFC/F658N & 10257 & 2004 Jul--2005 Jun & 15700 s\\
\hline
\end{tabular}
\end{center}
\end{table*}

\XMM\ observation 112220101 lasted 40 ks, collecting usable exposures on the quiescent NS of 33.5 ks with the pn, and 39.5 ks with the MOS1 and MOS2 detectors, all in full-frame mode with the medium filter.  
We reduced the data using SAS 13.0,\footnote{The \textit{XMM-Newton} SAS is developed and maintained by the Science Operations Centre at the European Space Astronomy Centre and the Survey Science Centre at the University of Leicester.} running {\tt emchain} and {\tt epchain} and filtering the event lists to retain predefined patterns 0$-$12 and use the \#XMMEA\_EM screening flags for MOS data, and retain patterns 0$-$4 and use the \#XMMEA\_EP flags for pn data.  
We used an extraction region of 19'' (as determined by Guillot et al. 2013 to maximize the signal-to-noise ratio), and extracted background from a nearby region on the same chip.  The background was low and fairly stable for the entire observation, so we used the entire valid exposure time.  

We reprocessed the \Chandra\ observations using CIAO v.4.5 \citep{Fruscione06}, correcting the data for charge transfer efficiency, and without using very-faint mode cleaning to reduce the background (as this can remove real data from sources of moderate brightness).
We extracted spectra from a 4'' radius extraction region (to deal with the relatively large point-spread function at the 4.4' off-axis angle), and a nearby, larger background region, using the {\tt specextract} CIAO script.  This includes the {\tt arfcorr} task that applies an energy-dependent correction of the effective area for the fraction of the point-spread function extracted.  

We combined the \Chandra\ spectra taken at the same epochs, combined the two MOS spectra, and grouped each spectrum to at least 20 counts/bin to permit the use of $\chi^2$ statistics.  No energy bin has a majority of counts outside the calibrated \Chandra\ energy range (0.277--9.886 keV).
We ignored spectral bins below 0.2 keV for \XMM\ data.  

\subsection{NGC 6397: \Chandra\ Data}\label{sec:6397}

There are five available \Chandra\ observations of NGC 6397 (Table 1); one 48 ks ACIS-I observation taken in 2000 \citep{Grindlay01b}, and pairs of ACIS-S observations in 2002 (totaling 55 ks) and in 2007 (totalling 237 ks; \citealt{Bogdanov10}). All of these observations were analyzed by \citet{Guillot11}.
We reprocessed the \Chandra\ observations using CIAO v.4.5, correcting the data for charge transfer efficiency, and without using the very-faint mode cleaning to remove background. We extracted spectra from a 2.5'' radius extraction region, and a nearby, larger background region, using the {\tt specextract} CIAO script, including energy-dependent correction of the effective area for the fraction of the point-spread function extracted. We grouped each spectrum to at least 20 counts/bin, except ObsID 7460 of NGC 6397 (the highest-quality spectrum, with 3124 counts) which we binned to at least 40 counts/bin.  We verified that different choices of binning do not significantly alter the results presented here.

\subsection{NGC 6397: HST Data}

Our dataset is described in \citet{Strickler09} and \citet{Cohn10}; we
provide a concise synopsis here (summarized in Table 1).  We used the \HST\ GO-10257 data set
(PI: Anderson), consisting of deep, highly dithered ACS/WFC images
targeting the center of NGC 6397 in F435W ($B$), F625W ($R$), and
F658N (H$\alpha$), over 10 single-orbit epochs, roughly once per month
between 2004 July and 2005 June.  The data include 5 short $B$ (13 s),
5 long $B$ (340 s), 5 short $R$ (10 s), 5 long $R$ (340 s), and 40
H$\alpha$ (390 s or 395 s) exposures, where the short exposures fill in
photometry for stars that saturate in the long exposures.  The large
number of H$\alpha$ frames produces a stacked H$\alpha$ image with
exceptionally good PSF sampling.

\section{NGC 6397: HST Analysis}\label{sec:6397hst}

\citet{Cohn10} report an analysis of the \HST\ dataset described above,
searching for optical counterparts for \Chandra\ X-ray sources.  The
star finding and photometry for this analysis used the software
described in \citet{Anderson08a}, developed for the ACS Globular
Cluster Treasury project.  Briefly, this software searches for pixels
that are higher than their eight neighbours (local peaks) in each
separate exposure, and identifies real stars at locations which are
peaks in more than a specified threshold fraction of the exposures.
It then measures the brightness of each star in each exposure, both
individually and simultaneously, using a spatially variable library of
PSFs.  This analysis did not detect any star within the error circle
of U24, the qLMXB in NGC 6397, although there are two relatively
bright stars only $\sim$0.6'' away from the center of the error circle, which has
 radius 0.285'' (at 95\% confidence; \citealt{Bogdanov10}).  However,
\citet{Cohn10} did note a ``hint of detection in $R$ only'' for U24 in
their Table 1 owing to the presence of ``a small `blip' near the
center of the error circle in the stacked $R$ image.''  They suggested
that it ``likely represents the combination of Airy ring artifacts''
from the two nearby bright stars.

\begin{figure*}
\begin{center}
\includegraphics[scale=0.8]{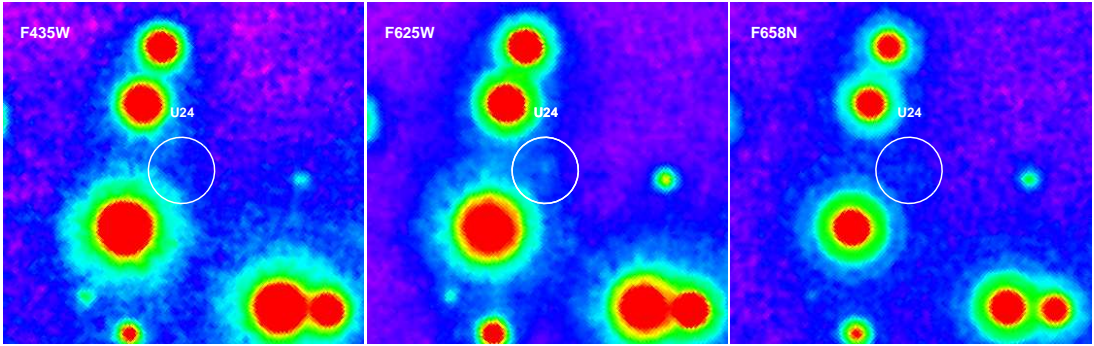}
\caption{Combined, oversampled, \emph{astrodrizzled} \HST\ ACS/WFC
  images of the region surrounding U24 in NGC 6397.  Left: combined
  $B$ image. Middle: combined $R$ image.  Right: combined H$\alpha$
  image. A possible star, and potential counterpart to U24, may be
  identified at the center of the 0.285'' error circle in the $R$ image, but
  is not present in the other frames.}
\label{HST images}
\end{center}
\end{figure*}

\begin{figure*}
\begin{center}
\includegraphics[scale=0.4]{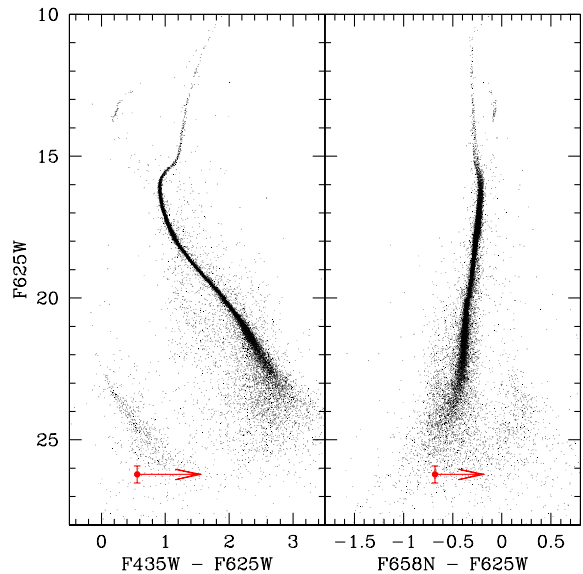} 
\caption{Left: \HST\ ACS/WFC color-magnitude diagram of NGC 6397 in the
  $B$ and $R$ filters.  The possible U24 counterpart is represented by
  an error bar with a horizontal arrow indicating the limit on its
  $B\!-\!R$ color based on its nondetection in $B$.  Right: \HST\
  ACS/WFC color-magnitude diagram of NGC 6397 in the H$\alpha$ and $R$
  filters.  The possible U24 counterpart is again represented by an
  error bar with a horizontal arrow indicating the limit on its
  H$\alpha$ excess based on its nondetection in H$\alpha$.}
\label{HST CMD}
\end{center}
\end{figure*}

We have conducted new photometry directly on the stacked images of
this \HST\ dataset, in order to assess the possibility that the
detection of a U24 counterpart in $R$ is real.  We utilized two
different methods for the stacking, the STSDAS \emph{astrodrizzle}
software (based on the drizzle algorithm, \citealt{Fruchter02}), and
stacking using the method described in \citet{Anderson08a}.  Both
methods involve oversampling the images by a factor of two.  The
stacked images (using \emph{astrodrizzle}) are presented in Figure 1.
Inspection reveals evidence for a faint star at the center of the
error circle in the $R$ image; this was evident with either stacking
approach.  Although the possible object is in a noisy region of the
image due to the wings of the PSF from the two much brighter stars,
the image suggests that this is a real, albeit weak, detection.  By
careful aperture photometry, we estimate an $R$ magnitude of
26.2$\pm$0.3 for this star (setting our zeropoints to those of
\citealt{Strickler09} and estimating the uncertainty from the
systematic error in determining the sky background).  At NGC 6397's
distance and extinction, this corresponds to $M_R$=13.7, which
corresponds to a maximum mass for the companion star (if a
main-sequence star) of  $\simle$0.089 \Msun\ \citep{Baraffe97}, and 
maximum $T_{eff}$=2900 K.

We do not find evidence for a faint star at the location of the
$R$-band candidate counterpart to U24 in the $B$ or H$\alpha$ images
(see Fig.~1).  In order to place an upper limit on the possible flux
enhancements in these bands, artificial stars with a range of
magnitudes were placed at this position.  The faintest magnitude that
produced a discernible flux enhancement was taken as representing the
upper flux limit.  This resulted in limits in $B>26.8$ and
H$\alpha>25.5$.  Thus, we find that $B\!-\!R>0.6$ and
H$\alpha\!-\!R>-0.7$.  We have plotted these photometric limits on the
CMDs reported by \citet{Cohn10} in Fig.~2.  Examination of these CMDs
indicates that the possible counterpart to U24 is consistent with
being no bluer than about the white dwarf sequence colour at this
magnitude and with showing no evidence for an H$\alpha$ excess
relative to the main sequence.  

The main sequence is difficult to identify at these magnitudes.  
However, comparison with the ultra-deep \HST\ imaging of NGC 6397
by \citet{Richer08} shows that the disappearance of the clear main sequence at $R\sim$25
 is largely due to the main sequence luminosity function dropping off, as luminosity falls off 
 dramatically with decreasing mass close to the hydrogen-burning limit.  
 The lack of a well-defined main sequence forces us to calculate the H$\alpha\!-\!R$ colour 
 of main-sequence stars at this position.  We use SYNPHOT to compute the 
 H$\alpha\!-\!R$ colour of a Pickles M5V 2950 K stellar model, using the appropriate extinction 
 and ACS filters, finding H$\alpha\!-\!R$=-0.75, which is consistent with the general trend of the  main sequence at brighter magnitudes. 
We can thus estimate the H$\alpha$ excess relative to the main
sequence, $\Delta \mbox{H}\alpha$, should not be larger than -0.25
mags (accounting for the uncertainty on our $R$ magnitude).  
We can then estimate a limit on the equivalent width (EW) of
H$\alpha$ in U24, using
EW(H$\alpha$)=RW$\times$[1-10$^{-0.4\times\Delta\mbox{H}\alpha}]$
\citep{DeMarchi10}, where RW is the rectangular width of the ACS/WFC
H$\alpha$ filter in \AA, i.e.\ 75 \AA.  This gives us an upper limit
of 19 \AA\ on the EW of H$\alpha$ for the putative U24
counterpart.

We can use this information to consider the possible nature of U24.  Quiescent LMXBs generally show strong H$\alpha$ emission, much stronger than during outbursts (when the optical emission is dominated by reprocessing of X-rays).  \citet{Fender09} show that typical EWs for H$\alpha$ of black hole LMXBs in quiescence are 30$-$300 \AA, but they have only three literature datapoints for NS LMXBs in quiescence, so we increase their sample here.  
\citet{Shahbaz96} found Cen X-4 to show an H$\alpha$ EW of 48 \AA, \citet{Torres02} an EW of 35$\pm$7 \AA, and \citet{D'Avanzo05} an EW of 40\AA. 
\citet{Garcia99} found 9.5\AA\ for Aql X-1 in quiescence, while \citet{Shahbaz96} found 4\AA, but all quiescent spectra of Aql X-1 before 1999 refer to the combined spectrum of Aql X-1 with its brighter, eastern neighbour 0.48" away \citep{Chevalier99}.  The companion-subtracted quiescent spectrum of Aquila X-1 shown in \citet{Chevalier99} suggests an EW of 30-40 \AA.
\citet{Casares02} show that XTE J2123-058 shows an EW of 20 \AA\ in H$\alpha$ in quiescence. 
\citet{Campana04b} find a 10 \AA\ EW for SAX J1808.4-3658, but this spectrum combines SAX J1808.4-3658 with a neighbour of similar brightness 0.5" away \citep{Hartman08,Deloye08}, so we may estimate (taking the continuum flux to be halved) a $\simge$20 \AA\ EW for SAX J1808.4-3658.
\citet{Bassa09} measure a 31 \AA\ H$\alpha$  EW for EXO 0748-676 in quiescence.
 The qLMXB in $\omega$ Cen exhibited a $\sim$1 magnitude H$\alpha$ excess \citep{Haggard04}, giving an estimated $\sim$110 \AA\ EW.
\citet{Beccari14} use H$\alpha$, $V$ and $I$ photometry to calculate H$\alpha$ EWs of 28 \AA\ and 50 \AA\ for the quiescent NS LMXBs W125 and W58/X5 in the globular cluster 47 Tuc. 
Finally, three systems (PSR J1023+0038, M28I=IGR J18245-2452, and XSS J12270-4859) swing between radio pulsations and active accretion \citep{Archibald09,Papitto13,Bassa14}. These three systems show no evidence of H$\alpha$ emission (or other disc signatures) when in their radio pulsar state \citep{Thorstensen05,Pallanca13,Bassa14}, but show H$\alpha$ EWs of 14-19 \AA\  \citep{Szkody03,Halpern13}, 72$\pm$5 \AA \citep{Pallanca13}, and 10-20 \AA\  \citep{Masetti06,Pretorius09}, respectively, when pulsations stop, and accretion, or a pulsar wind shock, produces low-level X-ray emission ($L_X\sim10^{33}$ ergs/s).  

What we take from this overview of the literature is that most
quiescent NS LMXBs show H$\alpha$ EWs typically between 20 to 50
\AA. The exception is the NS LMXBs which have swung between accretion
and radio pulsar activity, which have no H$\alpha$ emission during
their radio pulsar phase, and can have lower H$\alpha$ emission during
their near-quiescent periods.  As U24 in NGC 6397 shows no evidence of
active accretion (see \S 4.2 on X-ray variability below),
and has not been observed to be a radio pulsar despite numerous deep
pulsar searches of NGC 6397, it seems unlikely that U24 falls into the
latter category.  We should therefore expect an H$\alpha$ EW between
20$-$50 \AA, if hydrogen is present in the accreting material.  
Our upper limit on the H$\alpha$ EW of 19 \AA\ 
suggests, therefore, that  U24 does not possess hydrogen; and that the companion is a white dwarf.  
This interpretation is consistent with the weak limit
  that we are able to place on the $B\!-\!R$ color, which allows an
  object as blue as a white dwarf (much of the observed light may also come from an accretion disk).



\section{X-ray Spectral Fitting}\label{sec:result}

\subsection{$\omega$ Cen}

We begin by checking that we get similar results to \citet{Guillot13}, their Table 4 (lower part) and Fig. 6, when using the same data (though with slightly different processing and binning) and the same assumptions.  We use a {\it wabs*nsatmos} fit to only the \XMM\ pn spectrum and the (combined) 2000 \Chandra\ spectrum, and fix the distance and $N_H$ to Guillot et al's values ($d$=4.8 kpc, $N_H$=$1.82\times10^{21}$ cm$^{-2}$). For a fixed NS mass of 1.4 \Msun, we find a best-fit NS radius of 20.6 km, with 90\% confidence range of 18.1$-$23.5 km, in good agreement with \citet{Guillot13}. We list this fit as ``Guillot+13'' in Table 2.
(We use 90\% confidence ranges throughout, unless otherwise specified.)  A steppar plot calculating the permitted mass and radius range for this fit (not shown) is in good agreement with their Fig.\ 6.  Freeing $N_H$ and setting the distance to our preferred value of 5.3 kpc, we find similar results; for the NS mass fixed to 1.4 \Msun, the best fit is R=20.3 km, with a range of 13.6-29.4 km. 

We then check the effect of switching our ISM model to {\it tbabs}, using \citet{Wilms00} abundances and \citet{Verner96} cross-sections, keeping our other assumptions the same.  We find that for a fixed 1.4 \Msun\ mass and 5.3 kpc distance, the best-fit NS radius is 16.3 km, with range 11.3$-$23.0 km.  (The best-fit value of $N_H$ remains similar, at $1.8\times10^{21}$ cm$^{-2}$.)  
This is a $\sim$25\% change in the best-fit NS radius (or a 20\% change in the radius lower limit) due to a change in the ISM model, reinforcing the critical importance of accurate modeling of the ISM.  This effect seems to be due primarily to the reduction in the relative abundance of oxygen (compared to similar-Z elements) between the \citet{Anders82} and \citet{Wilms00} models, which leads to a shallower edge at 0.53 keV, for similar averaged interstellar absorption (see Fig.\ 3 for an illustration).  The downward curvature of the spectrum of the neutron star at low energies is due to a combination of the declining detector response, the temperature of the NS model (i.e. the intrinsic spectral curvature), and the effect of the ISM.  Thus, changing the shape of the ISM model alters how much of the observed curvature is attributed to intrinsic curvature of the spectrum, versus attributed to the effects of the ISM.

\begin{figure*}
\begin{center}
\includegraphics[scale=0.4,angle=270]{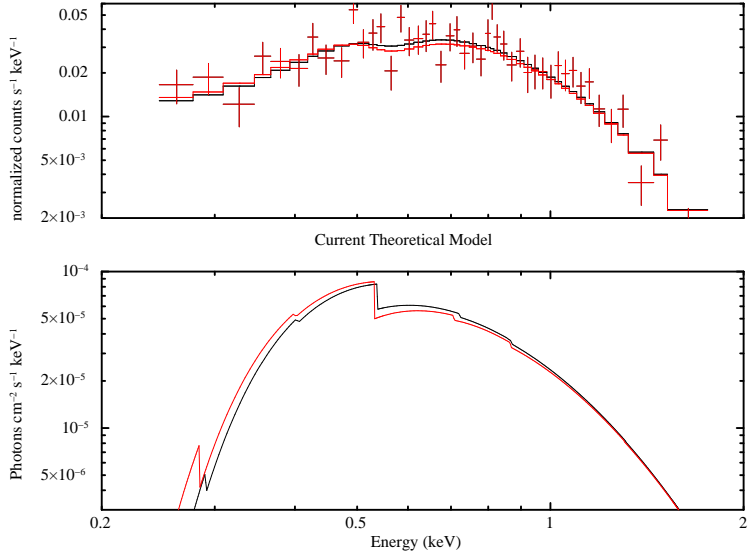} 
\caption{ Illustration of the effects of using the {\it wabs} versus {\it tbabs}  ISM models to fit the $\omega$ Cen qLMXB. The observed pn spectrum (crosses) and folded models (black, {\it tbabs} with {\it wilms} abundances; red, {\it wabs}) are plotted in the top panel, while the unfolded models are plotted in the bottom panel.  The {\it tbabs} model plotted has a higher $N_H$ ($1.6\times10^{21}$ cm$^{-2}$) than the {\it wabs} model ($1.3\times10^{21}$ cm$^{-2}$), to give similar total absorption, but the deeper 0.53 keV O edge of the {\it wabs} model produces a different shape to the absorption.
 }
\label{w Cen spectra}
\end{center}
\end{figure*}

For completeness, we check the effect of using the {\it phabs} model with XSPEC's default abundances \citep{Anders89} and cross-sections \citep{BalucinskaChurch92}, finding a best-fit radius of 24.4 (15.8 to $>$30) km, which is even farther from the results using {\it tbabs} and Wilms abundances (almost a 50\% difference) than when using the {\it wabs} model.  Details of each of these fits are listed in Table 2.

\begin{table*}
\begin{center}
\caption{\bf X-ray spectral fits to $\omega$ Cen; comparing to Guillot+13}
\begin{tabular}{llllll}
\hline
Fit & $N_H$                       & R & M            & $kT_{\rm eff}$ & $\chi^2$/dof \\
     & $\times10^{21}$ cm$^{-2}$ & km &  \Msun & eV &    \\
\hline
\multicolumn{6}{c}{\it wabs*nsatmos}\\
\hline
Guillot+13, d=4.8 & (1.82)         & 20.6$^{+2.9}_{-2.5}$ & (1.4) &  61$^{+4}_{-3}$ &  79.6/71 \\
d=5.3,$N_H$ free & 1.7$^{+0.5}_{-0.4}$ & 20.3$^{+9.1}_{-6.7}$ & (1.4) & 63$^{+11}_{-9}$ & 79.3/70 \\
\hline
\multicolumn{6}{c}{\it phabs*nsatmos}\\
\hline
d=5.3,phabs/angr & 1.9$^{+0.4}_{-0.5}$ & 24.4$^{+6^h}_{-8.5}$ & (1.4) & 60$^{+10}_{-4}$ & 80.0/70 \\
\hline
\multicolumn{6}{c}{\it tbabs*nsatmos}\\
\hline
d=5.3,wilm & 1.8$^{+0.5}_{-0.5}$ & 16.3$^{+6.7}_{-5.0}$ & (1.4) & 68$^{+12}_{-8}$ & 79.4/70 \\
\hline
\end{tabular}
\end{center}
\smallskip
Fits to only the \XMM/pn and 2000 \Chandraacis\ data on the $\omega$ Cen qLMXB.  Parameters in parentheses are fixed, others show 90\% confidence errors on a single parameter.  $^h$-parameter reached hard limit of model.
 The first fit matches a similar fit in \citet{Guillot13}.  See text for details of the assumptions in each fit.
\label{tab:matchGuillot_wCen}
\end{table*}

We next add in the combined MOS spectrum, and the new 2012 \Chandra\ data, so we are now fitting four spectra.  A {\it wabs*nsatmos} model fit to all the data finds a substantially smaller best-fit radius, 11.5 km (8.0$-$14.9 km) for a 1.4 \Msun\ NS at 5.3 kpc.  
Switching to {\it tbabs}, with new abundances and cross-sections, shrinks the best-fit radius further, to 10.0 km (5.0$-$12.6 km).   We note that the fitted $N_H$, including the new data and recent ISM modeling, is more consistent with the predicted $N_H$ to the cluster, as suggested by \citet{Lattimer14} with the rationale that it would make the NS mass and radius predictions consistent with the other datasets.

\begin{table*}  
\begin{center}
\caption{\bf X-ray spectral fits to $\omega$ Cen, all data}
\begin{tabular}{lllllll}
\hline
Fit & $N_H$                       & R & M            & $kT_{\rm eff}$ & PL flux & $\chi^2$/dof \\
     & $\times10^{21}$ cm$^{-2}$ & km &  \Msun & eV & ergs/cm$^{-2}$/s &   \\
\hline
\multicolumn{7}{c}{\it wabs*nsatmos}\\
\hline
wabs & 1.2$^{+0.3}_{-0.3}$ & 11.5$^{+3.4}_{-3.5}$ & (1.4) & 80$^{+18}_{-9}$ & - & 152.7/130 \\
\hline
\multicolumn{7}{c}{\it tbabs*nsatmos}\\
\hline
wilms & 1.3$^{+0.3}_{-0.3}$ & 10.0$^{+2.6}_{-5.0^h}$ & (1.4) & 85$^{+61}_{-9}$ & - & 152.4/130 \\
\hline
\multicolumn{7}{c}{\it tbabs*(nsatmos+pegpwrlw)}\\
\hline
NS+PL,M=1.4,d=5.3 & 1.3$^{+0.3}_{-0.3}$  & 10.3$^{+3.2}_{-5.3^h}$  & (1.4) & 83$^{+55}_{-10}$ & $6^{+21}_{-6}\times10^{-16}$ & 152.2/129 \\
NS+PL,M free & 1.3$^{+0.4}_{-0.3}$ & 10.6$^{+2.4}_{-1.8}$@ & 0.5$^{+1.6}_{-0^h}$@ & 74$^{+44}_{-6}$ & $6^{+21}_{-6}\times10^{-16}$ & 152.2/127 \\
\hline
\end{tabular}
\end{center}
\smallskip
Fits to \XMM/pn, \XMM/MOS, 2000 and 2012 \Chandraacis\ data on the $\omega$ Cen qLMXB.  Parameters in parentheses are fixed, others show 90\% confidence errors on a single parameter.  $^h$-parameter reached hard limit of model.   @-for the errors on mass and radius reported in this fit, the other parameter (e.g. radius, if mass was varied) was held fixed at its best-fit value.
\label{tab:wCen_alldata}
\end{table*}

We notice a clear residual in the \XMM\ pn spectrum above 1.5 keV, suggestive of a nonthermal component to the NS spectrum, as commonly observed in qLMXBs.  We test the addition of a power-law component (we use the {\it pegpwrlw} model, where the normalization is proportional to the intrinsic flux) to the spectrum, with photon index fixed to 1.0, 1.5, or 2.0 (the typical range of the spectra observed).  Such a component is not detected with 90\% confidence for any index, but the best fit power-law fluxes for each choice are similar, $F_X$(0.5$-$10 keV)=6-8$\times10^{-16}$ ergs/cm$^2$/s, or a best-fit $L_X\sim2\times10^{30}$ ergs/s, at 90\% confidence $L_X<9\times10^{30}$ ergs/s.  Permitting a power-law component slightly alters the best-fit NS radius to 10.2 km (5$-$13.3 km), changing the best fit by less than 0.1 km if the selected photon index is altered to a different value.  We fix the photon index to 1.5 below, and report the details of this fit (with mass fixed to 1.4 \Msun) in Table 3.

\begin{figure*}
\begin{center}
\includegraphics[scale=0.4,angle=270]{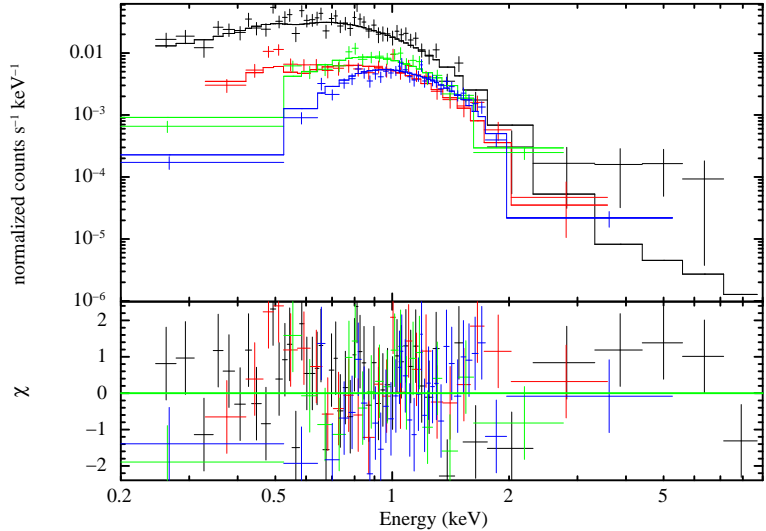} 
\caption{ Fit to spectra of $\omega$ Cen qLMXB, using {\it tbabs(nsatmos+pegpwrlw)} with index=1.5 and NS mass free.  Data and model are plotted in the top panel, with residuals plotted below.  Black: \XMM\ pn data; red: \XMM\ combined MOS data; green: \Chandra\ 2000 data; blue: \Chandra\ 2012 data.
 }
\label{w Cen spectra}
\end{center}
\end{figure*}

Finally, we create a {\it tbabs(nsatmos+pegpwrlw)} fit with index=1.5 and mass free.
This fit is illustrated in Figure 4, which shows all 4 spectra fit to the same model.
In the line for this fit in Table 3 (line labeled ``NS+PL,M free''), 
we provide the ranges for NS mass and radius if the other quantity is fixed at its best-fit value.  
In Figure 5 (left), we illustrate the full range of NS mass and radius values permitted by this fit, at 68\%, 90\%, and 99\% confidence ranges.  
The allowed values are consistent, assuming typical NS masses, with the most commonly discussed NS equations of state \citep[e.g.][]{Lattimer07}.
We then chose distances at the extremes of the reasonable distance range we identify above (5.13 and 5.47 kpc), and calculated probability contours for mass and radius for these distances (see Figure 5, right).

\begin{figure*}
\includegraphics[scale=0.4]{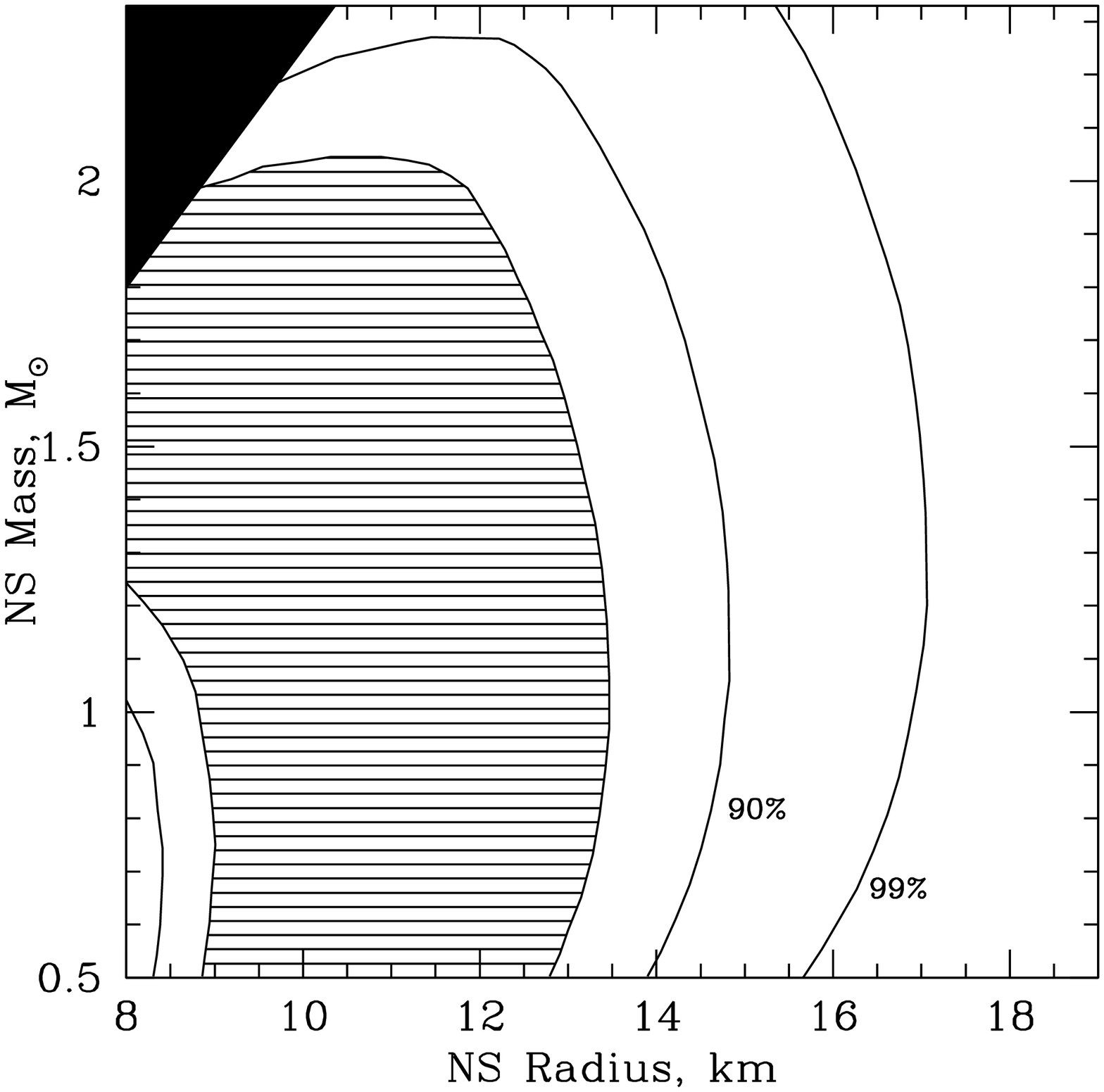} 
  \includegraphics[scale=0.4]{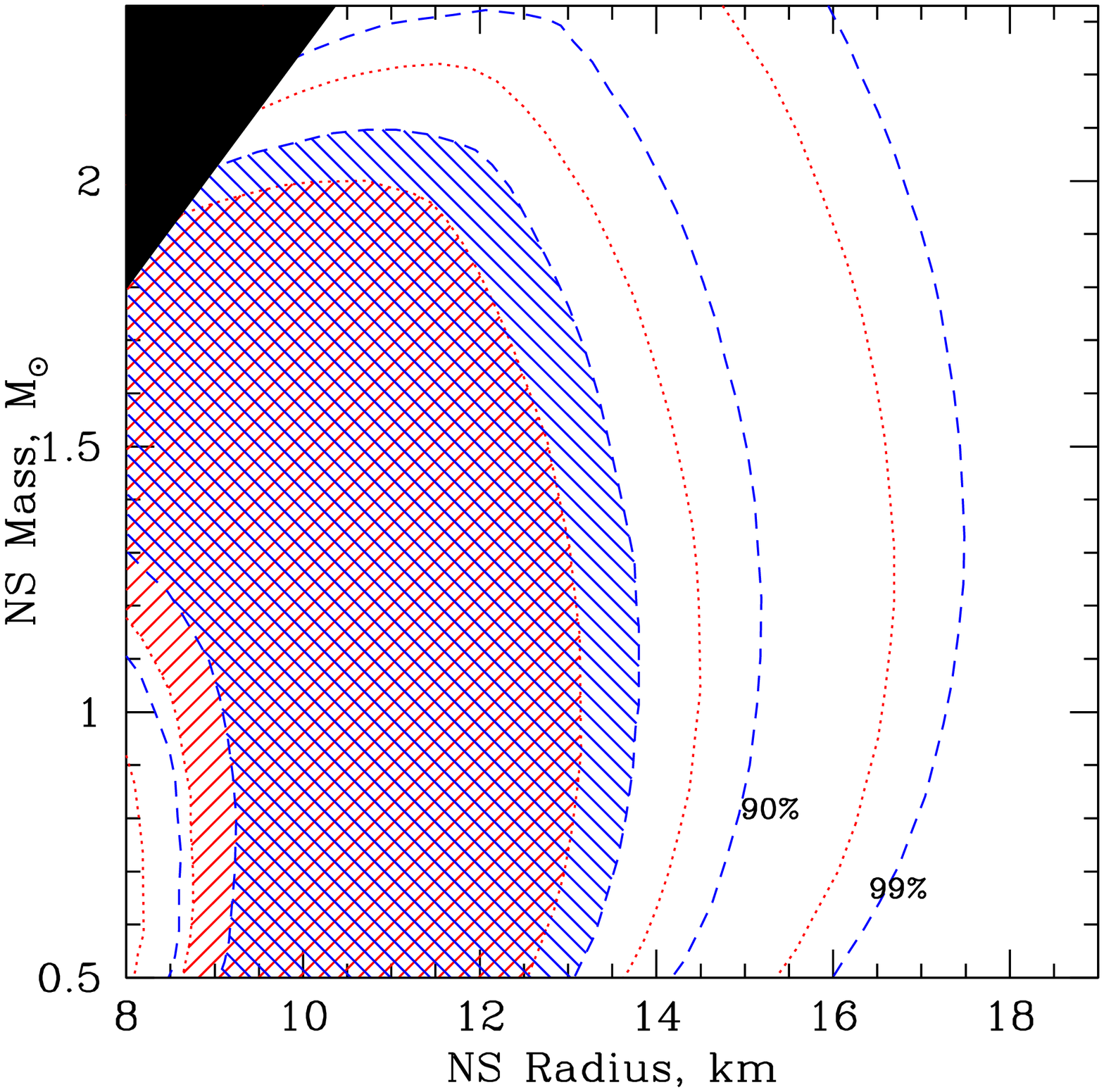}
 \caption{{\it Left:} Probability contours for the acceptable range of mass and radius for the $\omega$ Cen qLMXB.  This fit uses {\it tbabs(nsatmos+pegpwrlw)}, with assumed distance of 5.3 kpc and a powerlaw photon index of 1.5.  The solid black region at upper left identifies the excluded region (from causality constraints).  The shaded region is the 1$\sigma$ region, while the outer contours indicate the 90\% and 99\% confidence region.
{\it Right:} Probability contours for the $\omega$ Cen qLMXB, using the same assumptions as the left figure, except using assumed distances of 5.13 kpc (red, with $\nearrow$ shading and  dotted lines) or 5.47 kpc (blue, with $\searrow$ shading and dashed lines). }
\label{wCen contours}
\end{figure*}

We can use these data to ask whether this NS shows evidence of variability.  We use the two \Chandra\ spectra to measure any variation in the $\omega$ Cen NS temperature, between the years 2000 and 2012.  We use the {\it tbabs(nsatmos+pegpwrlw)} model, with d=5.3 kpc, photon index 1.5, and NS M=1.4 \Msun.  We first check whether permitting the power-law normalization to vary produces a better fit than fixing all parameters, but find that neither spectrum's power-law component varies from zero at 90\% confidence, and that an F-test indicates that untying the power-law normalization between the two spectra does not produce a better fit.  We therefore tie the power-law normalizations between the two spectra, and test whether varying the NS temperature produces a superior fit. Again, it does not (the F-test gives an F statistic of 0.657, probability of obtaining such a result by chance 0.42).  We freeze all parameters at their best-fit values, except for the 2000 \Chandra\ NS temperature, and measure a temperature difference of -0.7$^{+1.4}_{-1.4}$\%, at 90\% confidence. Alternatively, permitting a multiplicative constant between the models used for the two spectra, and holding all other parameters the same between the models, we find the 2012 observation to show a normalization 95$^{+5}_{-5}$\% of the 2000 observation.
 Thus, at 90\% confidence we can say that the $\omega$ Cen NS shows $<$2.1\% temperature variation, or $<$10\% flux variation, over 12 years.

Given the consistency of the flux over a decade, we can test how well the calibration of the \XMM\ and \Chandra\ instruments compare.  For this purpose, we used all the data, but separated them into three groups; the \Chandra, \XMM\ pn, and \XMM\ MOS spectra.  We froze all  parameters for the pn spectrum, and all parameters but the NS temperature for the others, and measured the difference in NS temperature recorded by the other instruments.  We found the MOS-measured temperature to be 1.6$^{+1.5}_{-1.7}$\% higher, and the \Chandra-measured temperature to be 0.7$^{+0.8}_{-0.8}$\% lower; thus, all instruments are consistent within the 90\% confidence errors.  Alternatively, starting from the best fit for a fixed NS mass of 1.4 \Msun, we fixed all parameters at their best-fit values, then freed the MOS and \Chandra\ radii.  Compared to the best-fit radius (fixed for the pn) of 10.50 km, the MOS result is 10.8$\pm0.3$ km, and the \Chandra\ result is 10.35$\pm0.15$ km, again consistent.  We infer that the relative uncertainties produced by differences between detectors contribute $\simle$4\% systematic uncertainties to NS radius measurements.  Of course, this does not account for possible absolute normalization uncertainties affecting all X-ray detectors, but the independent ground-based flux calibration procedures performed on the X-ray detectors \citep{Garmire03,Turner01,Struder01}  indicate that such uncertainties should be of the same order at most.



The value of the fitted $N_H$ has a very strong effect on the fitted radius of the NS.  Independently constraining the $N_H$ would thus significantly shrink the permitted region of mass and radius.  Is there another reliable method to constrain the $N_H$?  As described above, the extinction measured towards the cluster by CMD fitting gives the best estimate of the $N_H$ towards the cluster.  However, there are two caveats to using this; a) the conversion from optical extinction ($E(B-V)$) to X-ray measured $N_H$ is not perfectly measured or understood \citep[see][]{Guver09}; 
and b) there may be additional $N_H$ intrinsic to the system studied.  It might be possible to eliminate caveat a) by measuring the $N_H$ to other objects in the cluster with high precision.  We attempted this using spectra from the two brightest cataclysmic variables in $\omega$ Cen \citep{Carson00,Cool13}, extracted from the deep 2012 \Chandra\ observations.  However, we found that the measured $N_H$ for each spectrum was inconsistent with, and larger than, the value predicted by the extinction to the cluster, and the $N_H$ we measured for the qLMXB, and that the two $N_H$ values were also inconsistent with each other.  (Details of the spectral fitting of these sources will be presented elsewhere.)  Thus, we conclude that we cannot effectively constrain the $N_H$ to the $\omega$ Cen qLMXB by means other than spectral fitting.

We have shown above that the abundances used make a major difference in the interpretation of the spectra.  Here we illustrate this by taking our best spectral fit to the $\omega$ Cen data, changing the abundance pattern, and investigating the resulting allowed range in mass and radius.  The measured $N_H$ may be attributed either entirely to the interstellar medium (this may vary across the face of the cluster, and could show different abundances in different clouds), or to a mixture of interstellar $N_H$ and $N_H$ intrinsic to the binary.  
We test four fits using a single abundance pattern for the full $N_H$ column; {\it wilms} from \citet{Wilms00} (our preferred fit), {\it lodd} from \citet{Lodders03}, {\it aspl} from \citet{Asplund09}, and {\it angr} from \citet{Anders89}, where the first three are modern abundance measurements (which agree fairly closely), while the fourth is currently the default abundance pattern in XSPEC. 
We also consider a fit where we assume a fixed interstellar $N_H$ of 0.83$\times10^{21}$ cm$^{-2}$ (matching the measured $E(B-V)$ of 0.12 reported for $\omega$ Cen by \citealt{Harris96}, using the conversion of \citealt{Guver09}), with additional $N_H$ intrinsic to the binary.  To model additional $N_H$ intrinsic to the binary, we use the {\it zvfeabs} model in XSPEC, with the relative abundances of iron and other metals being set to 0.03 and 0.06 respectively (to simply represent the average $\omega$ Cen abundances of [Fe/H]=-1.5  \citealt{Harris96}, and [$\alpha$/Fe]=0.3 \citealt{Carney96}).

\begin{figure*}
\includegraphics[scale=0.4]{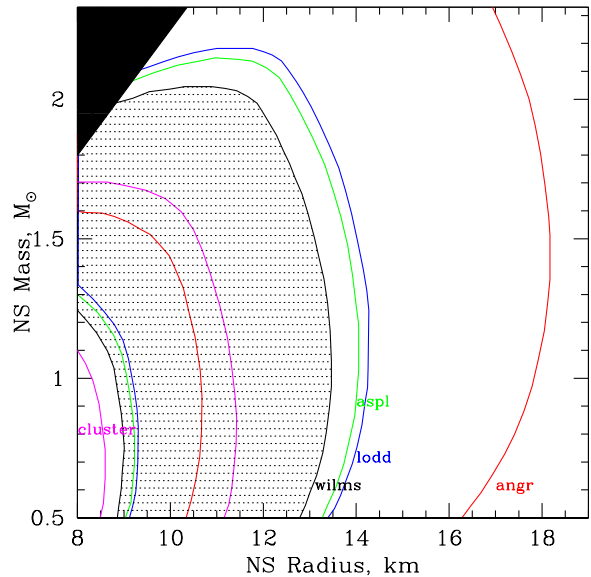} 
 \caption{ Five different 1$\sigma$ probability contours for the acceptable range of mass and radius for the $\omega$ Cen qLMXB, depending on the abundance pattern chosen.  This fit uses tbabs(nsatmos+pow), with assumed distance of 5.3 kpc and a powerlaw photon index of 1.5.  The solid black region at upper left identifies the excluded region (from causality constraints).  The shaded (dots) region is the 1$\sigma$ region for {\it wilms} abundances \citep{Wilms00}. The blue and green contours enclose the (similar) 1$\sigma$ regions for the {\it aspl} \citep{Asplund09} and {\it lodd} \citep{Lodders03} abundance choices, and the red contours enclose this region for the {\it angr} \citep{Anders89} abundances. The magenta contours are for a fixed column with {\it wilms} abundances, plus additional $N_H$ at the cluster abundance; see text.
}
\label{wCen abund}
\end{figure*}

 In Figure 6, we show the 1$\sigma$ confidence contours for these five abundance pattern choices.  Hearteningly, there is little difference between the contours for the three modern abundance measurements.  However, the \citet{Anders89} abundances make a significant difference (dramatically increasing the inferred radius, and generally making the contours wider), and we do not recommend their use.  Furthermore, we see a significant effect on the contours (toward smaller radii) when assuming that any $N_H$ above the reported cluster value has the abundances of the cluster.


\subsection{NGC 6397}
As above, we begin by checking that we get similar results to \citet{Guillot13}, their table 4 (lower part) and figure 6, when using the same data (though with slightly different processing and binning) and the same assumptions.  Using the {\it wabs*nsatmos} model, assuming a 2.02 kpc distance and $N_H=9.6\times10^{20}$ cm$^{-2}$, we find a best-fit mass, radius and temperature close to those of Guillot et al. (see Table 4).  Since Guillot et al. find a best-fit mass significantly different from 1.4 \Msun, we report {\it wabs*nsatmos} fits with and without a fixed 1.4 \Msun\ NS mass, noting that the fit using a 1.4 \Msun\ NS is strongly disfavored.  We also include a {\it wabs*nsatmos} fit changing the distance to our preferred 2.51 kpc value, and leaving the $N_H$ free, which enables a reasonable fit with a 1.4 \Msun\ NS, though the best-fit radius is uncomfortably low.

We then test the {\it tbabs*nsatmos} model, with \citet{Wilms00} abundances, and for completeness the {\it phabs*nsatmos} model using default \citep{Anders89} abundances.  In contrast to our $\omega$ Cen fits, we do not find dramatic discrepancies here between the different $N_H$ models (Table 4), perhaps due to the different effective area energy dependences of the \XMM\ EPIC vs. \Chandra/ACIS detectors.  If we add a power-law component, we find a significant improvement in the $\chi^2$ (with probability $2\times10^{-4}$ of occurring by chance, according to an F-test).  The power-law photon index $\Gamma$ finds a best fit at the (unphysical) value of 3.9, but is poorly constrained (range 0.5 to 4.7).  Varying the index in the range (1-2) observed for other quiescent NSs \citep[e.g.][]{Campana98a,Cackett10} makes little difference (e.g. the inferred radius upper limit changes by only 0.2 km); we include the fit with $\Gamma$ fixed to 2.0 below as an example.  Again, we see that an uncomfortably small radius ($<$9 km) is required for a 1.4 \Msun\ NS in all physically reasonable fits using a hydrogen atmosphere.  

We investigate the range of reasonable radii and masses for the last fit above, {\it tbabs(nsatmos+pegpwrlw)}, using a 2.51 kpc distance and including a power-law, with  photon index frozen to 1.5.  We plot the 1$\sigma$, 90\%, and 99\% confidence ranges in Figure 7 (left).  Clearly, the hydrogen-atmosphere fit indicates a NS smaller than 10 km, and suggests a mass below 1.4 \Msun.  
We do the same analysis for the extrema of our distance range, 2.44 and 2.58 kpc, and plot the results in Figure 7 (right).  

\begin{figure*}
\includegraphics[scale=0.4]{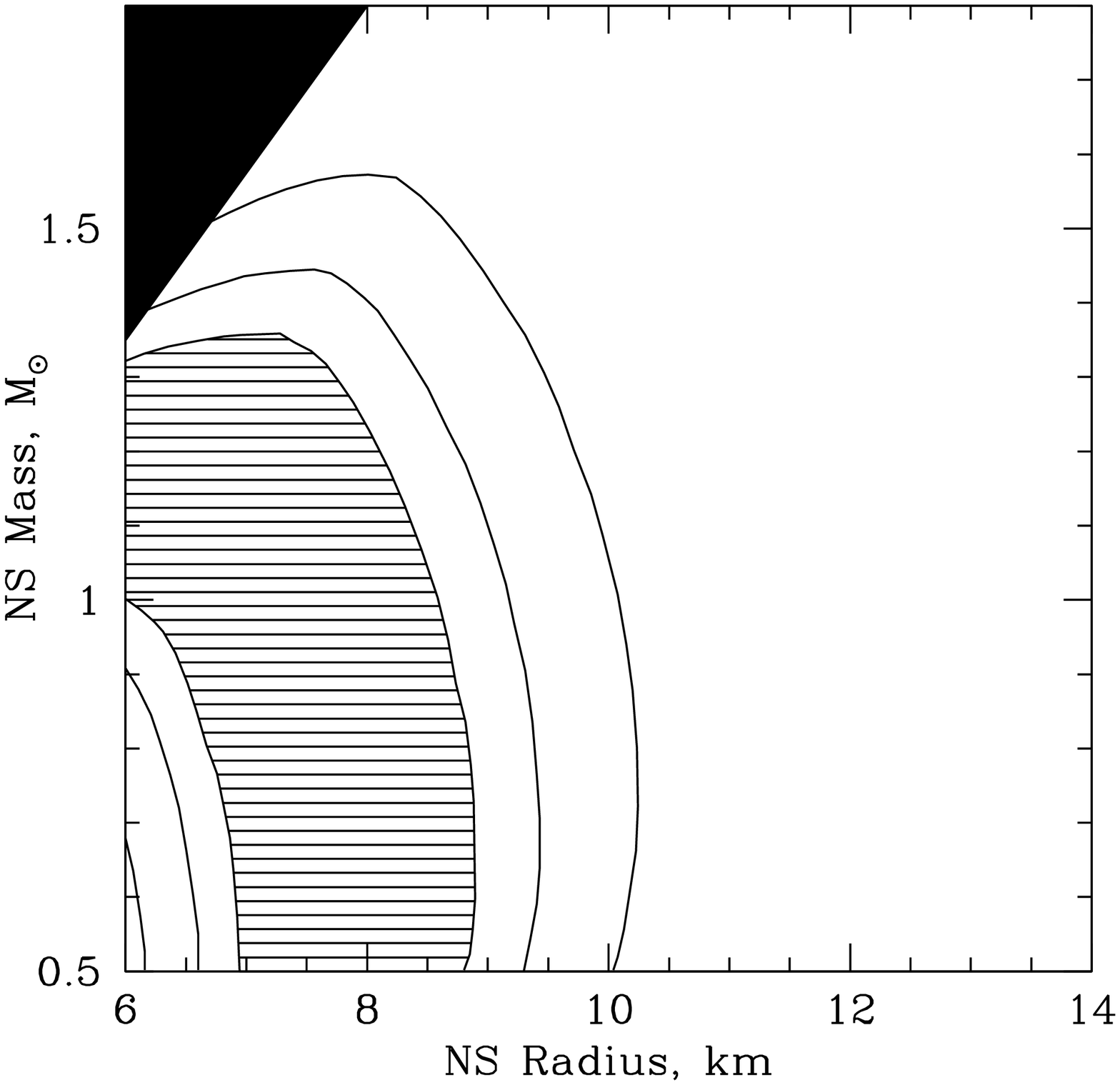} 
\includegraphics[scale=0.4]{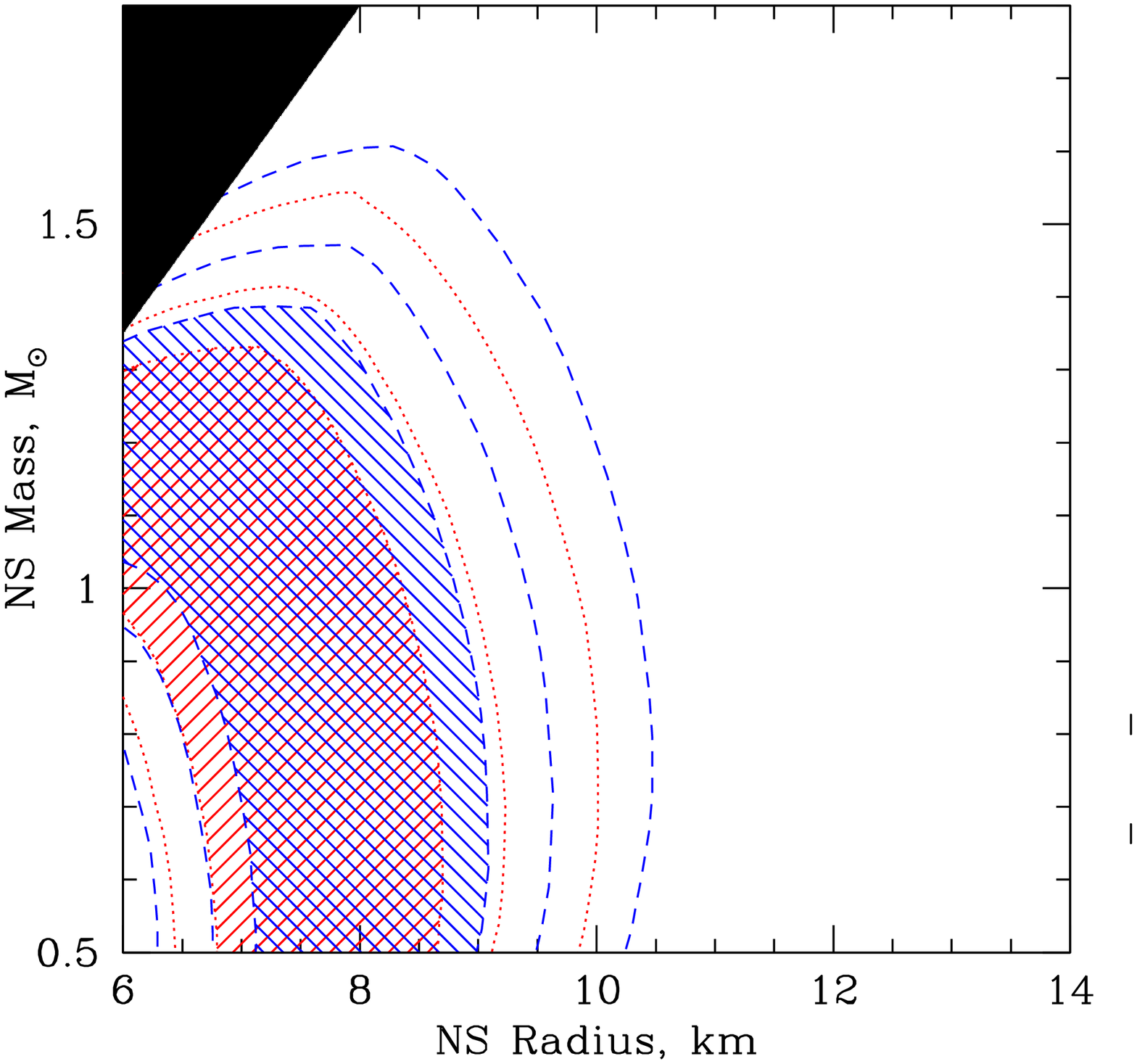}
\caption{Probability contours for the acceptable range of mass and radius for the NGC 6397 qLMXB, using a hydrogen atmosphere.  This fit uses tbabs(nsatmos+pow), with assumed distance of 2.51 kpc and a powerlaw photon index of 1.5.  The solid black region at upper left identifies an excluded region (from causality constraints).  The shaded region is the 1$\sigma$ confidence region, while the outer contours indicate the 90\% and 99\% confidence region.
{\it Right:} Probability contours for the NGC 6397 qLMXB, using the same assumptions as the left figure, except using assumed distances of 2.44 kpc (red, with $\nearrow$ shading and  dotted lines) or 2.58 kpc (blue, with $\searrow$ shading and dashed lines).   }
\label{6397 Hyd contours}
\end{figure*}

\begin{table*}
\begin{center}
\caption{\bf X-ray spectral fits to NGC 6397}
\begin{tabular}{lllllll}
\hline
Fit & $N_H$                       & R & M            & $kT_{\rm eff}$ & PL flux & $\chi^2$/dof \\
     & $\times10^{21}$ cm$^{-2}$ & km &  \Msun & eV & ergs/cm$^{-2}$/s &    \\
\hline
\multicolumn{6}{c}{\it wabs*nsatmos}\\
\hline
Guillot+13, d=2.02 & (0.96)  & 6.5$^{+1.0}_{-1.5^h}$ & 0.90$^{+0.21}_{-0.40^h}$ &  78$^{+10}_{-10}$ & - & 216.2/202 \\
Guillot+13, d=2.02 & (0.96)  & 6.9$^{+0.2}_{-0.3}$ & (1.4) &  87$^{+2}_{-3}$ & - &  277.6/203 \\ 
d=2.51,$N_H$ free & 1.0$^{+0.1}_{-0.1}$ & 7.4$^{+1.1}_{-1.4}$ & (1.4) & 89$^{+18}_{-8}$ & - & 219.9/202 \\
\hline
\multicolumn{6}{c}{\it phabs*nsatmos}\\
\hline
d=2.51,angr & 1.1$^{+0.1}_{-0.1}$ & 7.0$^{+1.9}_{-1.2}$ & (1.4) & 94$^{+19}_{-14}$ & - & 220.3/202 \\
\hline
\multicolumn{6}{c}{\it tbabs*nsatmos}\\
\hline
d=2.51,wilm & 1.3$^{+0.5}_{-0.5}$ & 7.3$^{+0.7}_{-1.1}$ & (1.4) & 89$^{+13}_{-6}$ & - & 226.9/202 \\
\hline
\multicolumn{6}{c}{\it tbabs(nsatmos+pegpwrlw)}\\
\hline
$\Gamma$=1.5 & 1.2$^{+0.1}_{-0.1}$ & 7.3$^{+1.1}_{-1.4}$  & (1.4) & 88$^{+19}_{-8}$ & $5^{+2}_{-2}\times10^{-15}$ & 211.8/201 \\
$\Gamma$=2.0 & 1.2$^{+0.1}_{-0.1}$ & 7.4$^{+1.1}_{-1.5}$  & (1.4) & 88$^{+21}_{-8}$ & $6^{+2}_{-2}\times10^{-15}$ & 211.4/201 \\
$\Gamma$=1.5,M free & 1.0$^{+0.2}_{-0.2}$ & 7.1$^{+1.4}_{-2.1^h}$@  & 1.07$^{+0.31}_{-1.07^h}$@ & 82$^{+43}_{-15}$ & $4^{+3}_{-3}\times10^{-15}$ & 208.5/200 \\
\hline
\multicolumn{6}{c}{\it tbabs(spHe)}\\
\hline
     & 1.1$^{+0.2}_{-0.2}$ & 7.5$^{+2.2}_{-2.3}$  & (1.4) & 86$^{+39}_{-13}$ & - & 213.1/202 \\
\hline
\multicolumn{6}{c}{\it tbabs(spHe+pegpwrlw)}\\
\hline
$\Gamma$=1.5 & 1.2$^{+0.2}_{-0.2}$ & 9.0$^{+2.9}_{-4^h}$  & (1.4) & 77$^{+55}_{-11}$ & $3^{+3}_{-2}\times10^{-15}$ & 209.1/201 \\
$\Gamma$=1.5,M free & 1.2$^{+0.2}_{-0.2}$ & 8.8$^{+2.9}_{-3.3}$@ & 1.45$^{+0.34}_{-0.44}$@ & 78$^{+32}_{-20}$ & $3^{+3}_{-2}\times10^{-15}$ & 209.0/200 \\  
\hline
\end{tabular}
\end{center}
\smallskip
Fits to the five \Chandraacis\ datasets on the NGC 6397 qLMXB.  Parameters in parentheses are fixed, others show 90\% confidence errors on a single parameter.  
 The first fit matches a similar fit in \citet{Guillot13}.  See text for details of the assumptions in each fit.  $^h$-parameter reached hard limit of model.  @-for the errors on mass and radius reported in this fit, the other parameter (e.g. radius, if mass was varied) was held fixed at its best-fit value.
\label{tab:6397}
\end{table*}

Considering the intriguingly small radius predicted by the NSATMOS hydrogen atmosphere model, and the evidence above (\S 3) suggesting an ultracompact nature for the NGC 6397 qLMXB, we consider a helium atmosphere model for this object.  Our helium atmosphere model, spHe, is described in \citet{Ho09}.\footnote{Now available as a local model in XSPEC, along with our carbon atmosphere models, in the NSX model package; see  https://heasarc.gsfc.nasa.gov/xanadu/xspec/models/nsx.html}  

\begin{figure*}
\begin{center}
\includegraphics[scale=0.4,angle=270]{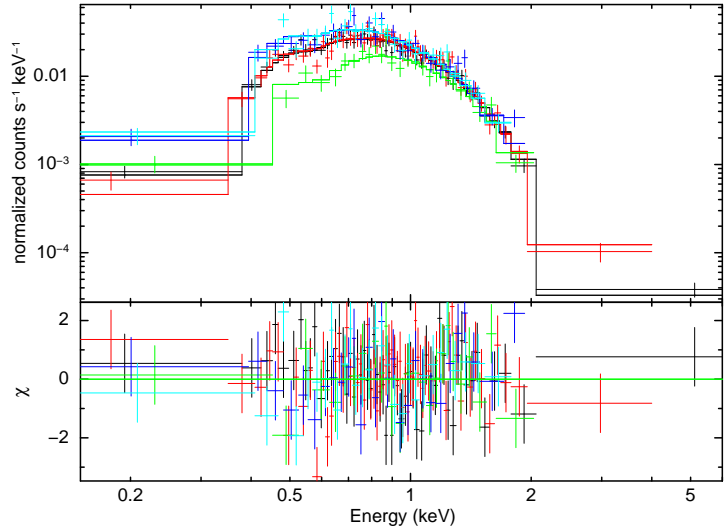} 
\caption{ Fit to spectra of NGC 6397 NS, using tbabs(spHe+pegpwr) with index=1.5 and NS mass free, with data and model plotted above, and residuals plotted below.  Black and red: \Chandra\ 2009 spectra; green: \Chandra\ 2000 data; light and dark blue: \Chandra\ 2002 data.
 }
\label{6397 spectra}
\end{center}
\end{figure*}

We first try fitting a {\it tbabs*spHe} model, without the power-law component.  This gives a relatively small radius (7.5$^{+2.2}_{-2.3}$ km for a 1.4 \Msun\ NS), and a good fit. 
 Adding a power-law component (with $\Gamma$ fixed to 1.5) improves the fit (the improvement is significant at 95\% confidence, according to an F-test).  This fit is shown in Figure 8. 
The availability of the power-law increases the allowed range of radii, giving a radius range of 9.0$^{+2.9}_{-4*}$ km (hitting the lower boundary of our model radius range).  
We find that our $N_H$ measurement is nicely consistent with the inferred $N_H$ from the extinction in the direction of NGC 6397 \citep{Harris96}, as again predicted by \citet{Lattimer14}.

\begin{figure*}
\includegraphics[scale=0.4]{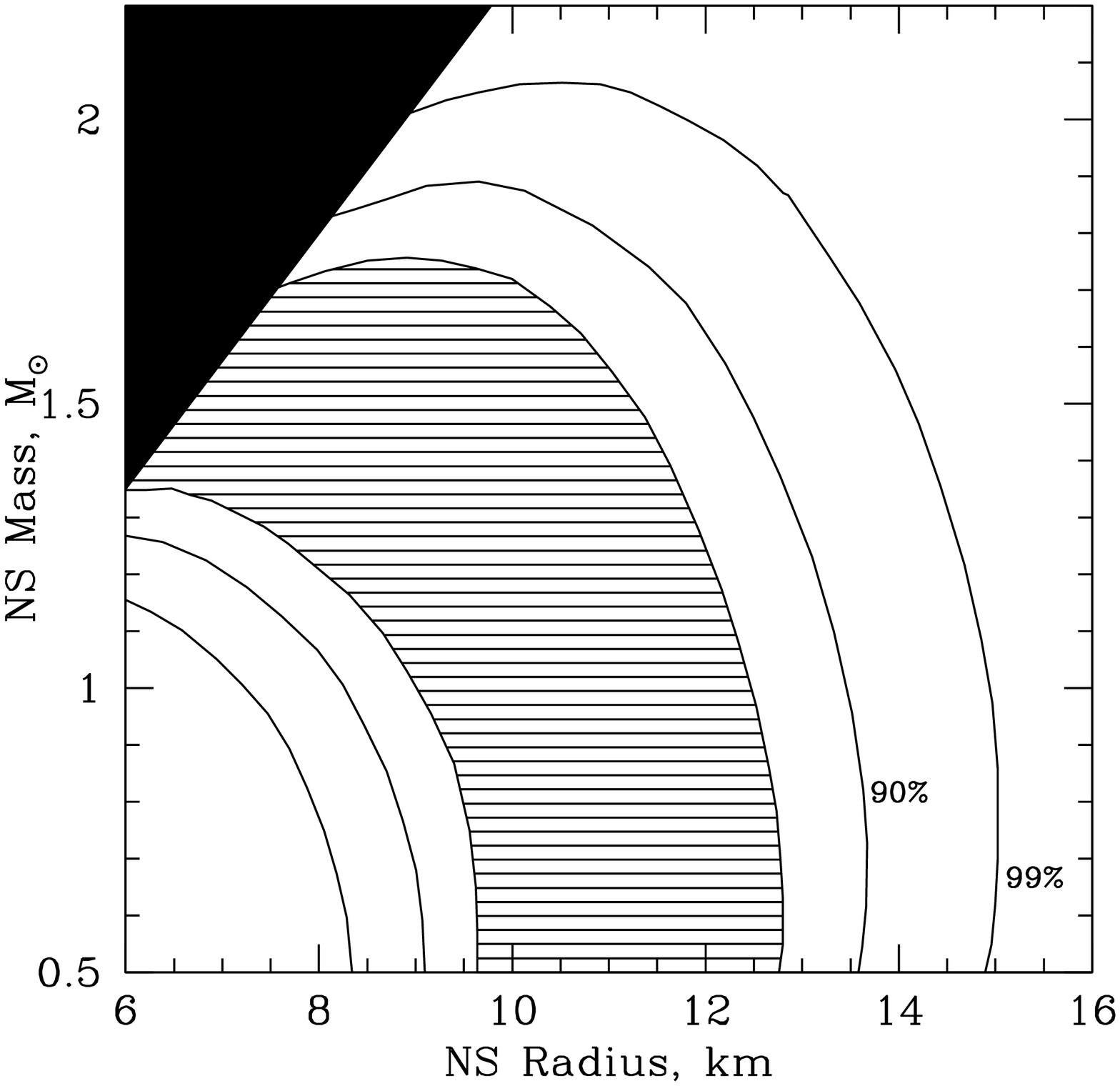} 
\includegraphics[scale=0.4]{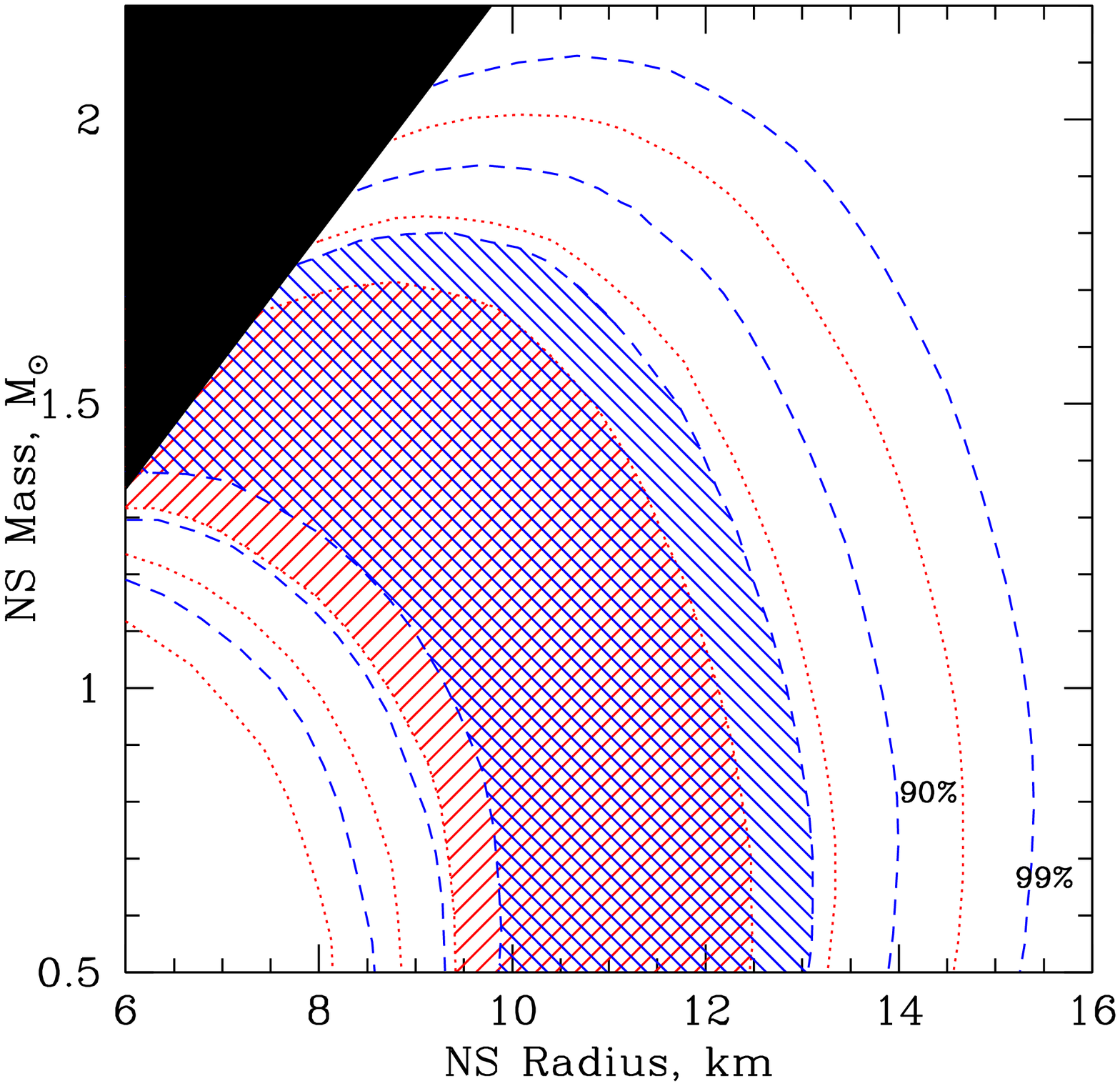}
\caption{Probability contours for the acceptable range of mass and radius for the NGC 6397 qLMXB, using a helium atmosphere.  This fit uses tbabs(spHe+pow), with assumed distance of 2.51 kpc and a powerlaw photon index of 1.5.  The black line at upper left identifies the excluded region (from causality constraints).  The shaded region is the 1$\sigma$ confidence region, while the outer contours indicate the 90\% and 99\% confidence region.
{\it Right:} Probability contours for the NGC 6397 qLMXB, using the same assumptions as the left figure, except using assumed distances of 2.44 kpc (red, with $\nearrow$ shading and  dotted lines) or 2.58 kpc (blue, with $\searrow$ shading and dashed lines).   }
\label{6397 Helium contours}
\end{figure*}

We investigate the range of reasonable radii and masses for this fit, {\it tbabs(spHe+pegpwrlw)}, using a 2.51 kpc distance and including the power-law with  photon index of 1.5.  We plot the 1$\sigma$, 90\%, and 99\% confidence ranges in Figure 9 (left).  
Again, we do the same analysis for the extrema of our distance range, 2.44 and 2.58 kpc, and plot the results in Figure 9 (right).  The ranges of radii and masses calculated for the helium atmosphere model are significantly larger than those for the hydrogen atmosphere.

As for the $\omega$ Cen qLMXB above, we can test for variation in temperature or total flux among the three \Chandra\ epochs.  Fixing the other parameters, and the 2009 best-fit temperature, at their best-fit values, we allowed the temperature of the NS in 2000 and 2002 to vary.  The NS temperature in 2000 was -0.2$^{+1.2}_{-1.2}$\% lower than in 2009, and that in 2002 was 0.2$^{+0.8}_{-0.9}$\% higher than in 2009.  Thus, we conclude that there is no evidence for variability among the observations, with temperature variations less than 1.4\% (at 90\% confidence) over 10 years.  \citet{Guillot11} performed a similar (slightly different) analysis, and our result agrees with theirs.  

\begin{figure*}
\includegraphics[scale=0.4]{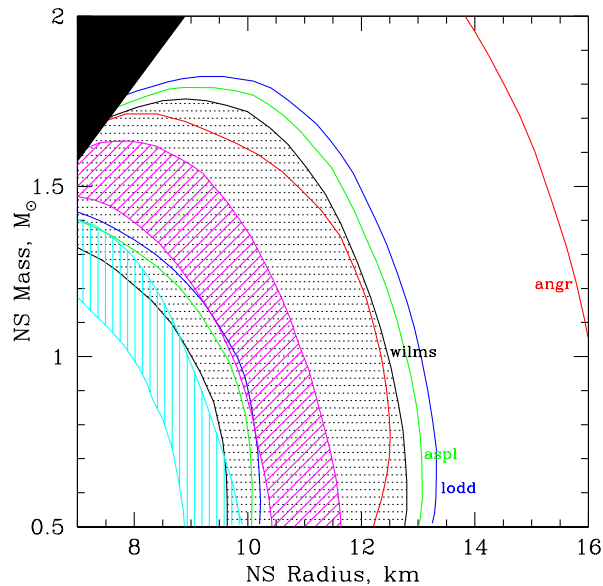} 
 \caption{ Six different 1$\sigma$ probability contours for the acceptable range of mass and radius for the NGC 6397 qLMXB, depending on the abundance pattern chosen.  This fit uses tbabs(spHe+pegpwrlw), with an assumed distance of 2.51 kpc and a powerlaw photon index of 1.5. 
The  region shaded with black dots is the 1$\sigma$ region for {\it wilms} abundances \citep{Wilms00}, as in Fig. 8. The blue and green contours enclose the (similar) 1$\sigma$ regions for the {\it aspl} \citep{Asplund09} and {\it lodd} \citep{Lodders03} abundance choices, and the red contours enclose this region for the {\it angr} \citep{Anders89} abundances. The magenta contours (shaded at 45$^{\circ}$ angle) are for a fixed column of $1.2\times10^{21}$ cm$^{-2}$ with {\it wilms} abundances, plus additional $N_H$ at the cluster abundance; the cyan contours (shaded vertically) are the same, except the fixed column is at $1.0\times10^{21}$ cm$^{-2}$.
}
\label{wCen contours}
\end{figure*}

Finally, we investigate the effect of different choices of abundance patterns, as for $\omega$ Cen, showing the results in Fig. 10.  We see, again, that the choice of the {\it wilms}, {\it aspl}, or {\it lodd} abundance patterns has only a small effect on the outcome.  The {\it angr} \citep{Anders89} abundance pattern again produces rather larger radii.  We show two different choices of a cluster $N_H$ column plus intrinsic absorption, with different selections of the cluster $N_H$ value.  Using the \citet{Predehl95} relation, the cluster $E(B-V)$ of 0.18 indicates $N_H$=$1.0\times10^{21}$ cm$^{-2}$, while the \citet{Guver09} relation indicates $N_H$=$1.2\times10^{21}$ cm$^{-2}$.  In each case, the allowed contours are smaller than when assuming the entire column is interstellar; the larger intrinsic absorption gives significantly smaller radii, though the $N_H$ difference is only $2\times10^{20}$ cm$^{-2}$.

\section{Discussion}

The two qLMXBs we analyze in this paper are the objects with the most extreme suggested mass and/or radius values in the analysis of \citet{Guillot13}.  In particular, the extremely low value of the inferred radius in \citet{Guillot13}'s analysis significantly reduced their final result.  Although \citet{Guillot13} argue that their targets show consistent radii within 2$\sigma$, it is clear that the primary result of their paper--a relatively low NS radius of 9.1$^{+1.3}_{-1.5}$ km--depends on their analysis of the NGC 6397 qLMXB.  Their simultaneous spectral fit omitting NGC 6397 finds a significantly larger NS radius of 10.7$^{+1.7}_{-1.4}$ km, which is fully consistent with the $\sim$11$-$13 km NS radius range predicted by nuclear experimental \citep[e.g.][]{Tsang12} and nuclear theoretical studies \citep[e.g.][]{Hebeler13}. 

For this reason, the consideration of a helium atmosphere for the NGC 6397 qLMXB is crucial for robust constraints on the NS radius.  
We have presented an optical detection of a likely candidate optical counterpart to the NGC 6397 qLMXB, and intriguing, though not conclusive, evidence in favor of a white dwarf companion, which would suggest the NS may have a helium atmosphere.    
Fitting the NGC 6397 qLMXB with hydrogen atmosphere models leads to an inferred radius $<$9.0 km at 90\% confidence, for any mass $>$1.2 \Msun, and within our considered distance range.  Such a low NS radius would significantly disagree with the nuclear theoretical and experimental studies cited above.  (Note that the lower distances inferred from dynamical distance estimates would exacerbate the discrepancy.)  We conclude that a helium atmosphere is preferred for the NGC 6397 qLMXB, making it the first such qLMXB for which we have any evidence of a helium atmosphere.

We find that the choice of ISM abundance model can make a significant difference in the inferred NS radius.  Furthermore, our best-fit $N_H$ values for both the $\omega$ Cen and NGC 6397 qLMXBs are closer to the $N_H$ inferred from the cluster optical extinction $E(B-V)$, using the \citet{Guver09} relation between $N_H$ and $E(B-V)$, than the values from the fits by \citet{Guillot13}.  For NGC 6397, our fitted $N_H$ value agrees with this extinction estimate ($1.2\pm0.2\times10^{21}$ cm$^{-2}$, vs. $1.2\times10^{21}$ cm$^{-2}$).  However, for $\omega$ Cen, our fitted $N_H$ value remains well above the cluster extinction estimate ($1.3\pm0.3\times10^{21}$ cm$^{-2}$, vs. $0.83\times10^{21}$ cm$^{-2}$), which does not account for small-scale variations.  We agree with \citet{Guillot13} that spectral fitting is the only reliable way to determine the $N_H$ to a particular X-ray source.  Considering the strong dependence of the inferred NS radius on details of the ISM abundance model, it is clear that the best targets for further constraints on the NS radius should have low $N_H$ values.

We find no evidence of variability among multiple observations of the $\omega$ Cen and NGC 6397 qLMXBs over decade timescales, constraining their temperature fluctuations to $<$2.1\% and $<$1.4\% respectively.  Combined with the strong constraints on variation of the M28 qLMXB \citep{Servillat12} and X7 in 47 Tuc \citep{Heinke06a}, we now have four qLMXBs that show little or no power-law component and with strong constraints on temperature variability over multi-year timescales.  This is evidence in favor of the suggestion \citep{Heinke03d} that the strength of the power-law component indicates the strength of any continuing accretion, and thus that the thermal emission from qLMXBs  showing no power-law component is powered entirely by deep crustal heating \citep{Brown98}.

We can ask the question of what the inferred orbital period of the NGC 6397 qLMXB might be, if it is ultracompact and its thermal emission is powered by deep crustal heating.  Using the ``standard cooling'' curve relating the mass transfer rate and bolometric NS luminosity of \citet{Yakovlev04}, and our calculated bolometric quiescent $L$ of $2.0\times10^{32}$ ergs/s, we estimate a mass transfer rate of $2.5\times10^{-12}$ \Msun/year.   Using the ultracompact X-ray binary evolution track of \citet{Deloye03}, we can thus predict an orbital period of 50 minutes for this system.  If the NS experiences enhanced neutrino cooling, then the mass transfer rate would be higher, and the orbital period could be lower, down to 21 minutes (the minimum for transient helium-accreting ultracompact X-ray binaries; \citealt{Lasota08}).  

Our constraints on the mass and radius of the $\omega$ Cen and NGC 6397 qLMXBs are not extremely constraining.  However, our results remove the evidence pointing towards a small NS radius advanced by \citet{Guillot13}.  \citet{Lattimer14} argued for different choices of $N_H$ and helium atmospheres for some of the five sources studied by \citet{Guillot13}.  Using analytical prescriptions to alter the inferred NS mass and radius ranges of Guillot et al. to match their choices, and including information about the behavior of neutron matter at low densities, and the NS maximum mass, they then calculated a larger inferred NS radius.  Our analysis can be taken as overall support for their arguments, as we provide explicit observational support for different choices of $N_H$ and for a helium atmosphere model, although the details are significantly different.

{\bf Acknowledgements}

We thank G.~R.~Sivakoff for helpful comments. 
 COH is supported by an NSERC Discovery Grant and an Alberta Ingenuity New Faculty Award.  DH and AMC are supported by \Chandra\ Award Numbers GO2-13057A and GO2-13057B issued by the CXO, which is operated by the Smithsonian Astrophysical Observatory for and on behalf of NASA under contract NAS8-03060. 
Based in part on observations with the NASA/ESA Hubble Space Telescope, obtained at the Space Telescope Science Institute, which is operated by the Association of Universities for Research in Astronomy, Inc., under NASA contract NAS5-26555. These observations are associated with proposal GO-10257.
Support for program GO-10257 was provided by NASA through a grant from the Space Telescope Science Institute, which is operated by the Association of Universities for Research in Astronomy, Inc., under NASA contract NAS 5-26555. 
A portion of these results were based on observations obtained with \textit{XMM-Newton}, an ESA science mission with instruments and contributions directly funded by ESA Member States and NASA. 
This research made use of data obtained through observations made by NASA's  \textit{Chandra X-ray Observatory}, data obtained from the \Chandra\ Data Archive, and data obtained from the High Energy Astrophysics Science Archive Research Center (HEASARC), provided by NASA's Goddard Space Flight Center. 
This research has made use of the NASA Astrophysics Data System (ADS), the electronic archive arXiv.org maintained by Cornell University Library, and software provided by the \Chandra\ X-ray Center (CXC) in the application package CIAO.

\bibliographystyle{mn2e}
\bibliography{odd_src_ref_list}

\end{document}